\documentclass[aps,prd,twocolumn,superscriptaddress,longbibliography,nofootinbib,10pt]{revtex4-2}
\usepackage{graphicx}
\usepackage[hidelinks]{hyperref}
\usepackage{amsmath}
\usepackage{amssymb}
\usepackage[dvipsnames]{xcolor}
\usepackage{orcidlink}
\usepackage{xspace}

\newcommand{\Fig}[1]{Fig.~\ref{#1}}
\newcommand{\Eq}[1]{Eq.~(\ref{#1})}
\newcommand{\Eqs}[2]{Eqs.~(\ref{#1}) and (\ref{#2})}
\newcommand{\Sec}[1]{Sec.~\ref{#1}}

\newcommand{\App}[1]{Appendix~\ref{#1}}
\newcommand{\mpl}{m_{\rm pl}}

\def\newacronym#1#2#3{\gdef#1{\gdef#1{#2\xspace}#3 (#2)\xspace}}
\newacronym{\FLRW}{FLRW}{Friedmann--Lema\^itre--Robertson--Walker}
\newacronym{\BSSN}{BSSN}{Baumgarte--Shapiro--Shibata--Nakamura}
\def\MPl{M_{\rm{Pl}}}

\usepackage{ifthen}
\newboolean{editorial}
\setboolean{editorial}{true}
\newcommand{\editorial}[2]{\ifthenelse{\boolean{editorial}}{\textcolor{red}{[\textsf{\textbf{{#1}}}: }\textcolor{blue}{\textsf{{#2}}}\textcolor{red}{]}}{}}

\begin{document}

\title{Cosmological First-Order Phase Transitions with Full Nonlinear Gravity}

\author{Campbell Benedix\orcidlink{0009-0002-9382-4422}}
\email{benedix1@kenyon.edu}
\affiliation{Department of Physics, Kenyon College, 201 N College Rd, Gambier, OH 43022, USA}

\author{Eve Currens\orcidlink{0000-0002-8519-040X}}
\email{ecurrens@umass.edu}
\affiliation{Department of Physics, Kenyon College, 201 N College Rd, Gambier, OH 43022, USA}
\affiliation{Department of Physics, University of Massachusetts Amherst, MA 01003, USA}

\author{John~T.~Giblin~Jr.\orcidlink{0000-0003-1505-8670}}
\email{giblinj@kenyon.edu}
\affiliation{Department of Physics, Kenyon College, 201 N College Rd, Gambier, OH 43022, USA}
\affiliation{Department of Physics/CERCA/Institute for the Science of Origins, Case Western Reserve University, Cleveland, OH 44106-7079, USA}
\affiliation{Center for Cosmology and AstroParticle Physics (CCAPP) and Department of Physics, The Ohio State University, Columbus, OH 43210, USA}

\author{Chloe~Richards\orcidlink{0000-0001-6935-6708}}
\email{richards4@kenyon.edu}
\affiliation{Department of Physics, Kenyon College, 201 N College Rd, Gambier, OH 43022, USA}
\affiliation{Center for Cosmology and AstroParticle Physics (CCAPP) and Department of Physics, The Ohio State University, Columbus, OH 43210, USA}

\begin{abstract}
Cosmological first-order phase transitions can contribute to the stochastic gravitational wave background. For decades, these first-order phase transitions have been studied using a Friedmann--Lema\^itre--Robertson--Walker background where the gravitational waves are treated as linear perturbations. Here, we present simulations of a first-order phase transition in the presence of fully nonlinear general relativity, using the BSSN formalism.  We simulate a universe that contains a scalar field alongside a relativistic radiation fluid. We focus on GUT-scale transitions in which bubbles nucleate near the horizon size. We describe a method to construct initial data that satisfies the Hamiltonian and momentum constraints, starting from a Coleman--de~Luccia-like bubble profile and adjusting the bubble radius.  We compute the gravitational wave spectrum directly from the trace-free extrinsic curvature and find a present-day peak amplitude of $\Omega_{\rm GW,0}h^2 \sim 5\times 10^{-11}$, consistent with earlier results that do not include nonlinear gravity. The peak frequency depends on the symmetry-breaking scale and could fall anywhere from the LISA band to GHz frequencies. At late times, the field fragments into long-lived oscillons, which push its effective equation of state toward that of matter.
\end{abstract}

\maketitle

\section{Introduction}
\label{sec:intro}

Gravitational wave astronomy has only begun to teach us about the Universe.  In the past decade or so, the direct detection of gravitational waves from compact-binary mergers from the LIGO--Virgo--KAGRA collaboration~\cite{LIGOScientific:2016aoc,LIGOScientific:2026wfs} and the strong evidence of a stochastic gravitational wave background (SGWB) from pulsar timing arrays (PTAs)~\cite{NANOGrav:2023gor,EPTA:2023fyk,Reardon:2023gzh,Xu:2023wog} provide inspiration that gravitational waves will open new windows to the Universe at new scales.  

Stochastic gravitational waves are of particular interest as they can be generated in the early Universe, providing a direct probe of physical processes that are opaque to electromagnetic radiation~\cite{Caprini:2018mtu}.  In most cosmological contexts, the energy scale of the Universe at the time of the process is directly connected to the frequency of the gravitational radiation today~\cite{Easther:2006gt,Giblin:2014gra}, which positions the SGWB to probe a huge range of energy scales.  Proposed observatories such as the Laser Interferometer Space Antenna (LISA)~\cite{LISA:2017pwj,LISA:2024hlh}, the Einstein Telescope~\cite{Punturo:2010zz,ET:2019dnz}, and MHz--GHz detectors~\cite{Aggarwal:2020olq} will probe many decades of this frequency space and support our understanding of what high-energy processes might have occurred.

In the recent past, there have been many proposed sources of stochastic gravitational waves~\cite{Grishchuk:1974ny,
Starobinsky:1979ty,
Khlebnikov:1997di,
Easther:2006gt,
Easther:2007vj,
Garcia-Bellido:2007nns,
Garcia-Bellido:2007fiu,
Dufaux:2007pt,
Adshead:2018doq,
Amin:2014eta,
Lozanov:2019jxc,
Vachaspati:1984gt,
Damour:2004kw,
Blanco-Pillado:2017oxo,
Ananda:2006af,
Baumann:2007zm} including those produced during first-order phase transitions (FOPTs)~\cite{Witten:1984rs,Hogan:1986qda}.
In these scenarios, the universe transitions from a metastable state to a true-vacuum state via the nucleation and collision of bubbles of true vacuum~\cite{Coleman:1977py,Callan:1977pt,Linde:1981zj}.  The latent heat liberated during this process is deposited into the dynamical scalar field and, when present, the coupled plasma~\cite{Steinhardt:1981ct,Espinosa:2010hh}.
While the electroweak phase transition is likely not first-order~\cite{Kajantie:1996mn}, beyond-the-Standard-Model (BSM) physics could easily contain one or more of these processes at scales up to and including the grand unified theory (GUT) scale~\cite{Mazumdar:2018dfl,Hindmarsh:2020hop,Athron:2023xlk}.  
BSM physics can also push these signals into the LISA band~\cite{Caprini:2015zlo,Caprini:2019egz,Caprini:2024hue} or help to explain observations from PTA experiments~\cite{NANOGrav:2023hvm}, although this interpretation has been disputed~\cite{Wright:2024awr}.
FOPTs with varied equation of state parameters have also been of recent interest, both to understand their end-state solutions and to study their relevant energy scales and impact on gravitational wave production~\cite{Giombi:2024kju,Barni:2026ecg}.

Gravitational waves generated during a FOPT can be sourced by the gradients of the bubble walls themselves~\cite{Kosowsky:1991ua,Kosowsky:1992rz,Kosowsky:1992vn,Kamionkowski:1993fg,Huber:2008hg,Child:2012qg,Cutting:2018tjt} as well as from sound waves in the coupled plasma~\cite{Hindmarsh:2013xza,Giblin:2013kea,Giblin:2014qia,Hindmarsh:2015qta,Hindmarsh:2017gnf} and from magnetohydrodynamic turbulence~\cite{Kamionkowski:1993fg,Caprini:2015zlo}.  
In these studies, numerical simulations have been critical in understanding the nonlinear physics of the sources~\cite{Hindmarsh:2013xza,Giblin:2013kea,Giblin:2014qia,Hindmarsh:2015qta,Hindmarsh:2017gnf,Cutting:2018tjt,Cutting:2019zws,Jinno:2020eqg,Li:2026hef}.  
In almost all work to date, numerical implementations have relied on a homogeneous and isotropic \FLRW cosmological background on which the nonlinear dynamics of the constituent parts of the Universe live.  The inhomogeneity of the fields and fluids then sources tensor perturbations on top of this background. 
This ends up being a consistent assumption when it is clear that the Einstein equations are trusted at linear order: for example, when bubbles are small compared to the Hubble horizon and/or the density contrasts are sufficiently small.  At the same time, there could be FOPTs that are stronger or slower~\cite{Ellis:2019oqb,Cutting:2019zws}, which may increase the size of the density contrast or move the bubble size closer to the Hubble scale \cite{Jinno:2023vnr, Jinno:2024nwb}.  
If strong gravity is important during these transitions, e.g.~\cite{Jung:2021mku, Giombi:2023jqq}, 
fully nonlinear simulations will be required to fully resolve the physical and observational consequences of the transition.  Beyond that, nonlinear gravity also includes all orders of perturbation theory and can help us understand whether linear-order perturbation theory is sufficient to calculate gravitational wave backgrounds, or not~\cite{Bastero-Gil:2010tpb}.

Fully nonlinear gravity is used in cosmological contexts to study late-time cosmological inhomogeneities~\cite{Giblin:2015vwq,Bentivegna:2015flc,Macpherson:2016ict,Giblin:2016mjp,East:2017qmk}, the robustness of inflation to inhomogeneous initial conditions~\cite{East:2015ggf,Clough:2016ymm,Clough:2017efm,Aurrekoetxea:2019fhr,Joana:2020rxm,Corman:2022alv,Elley:2024alx}, cosmic bubble collisions~\cite{Johnson:2011wt,Wainwright:2013lea,Wainwright:2014pta,Johnson:2015gma}, preheating and oscillon formation~\cite{Giblin:2019nuv,Kou:2019bbc,Kou:2021bij,Aurrekoetxea:2023jwd,Nazari:2020fmk}, cosmic string loops~\cite{Helfer:2018qgv,Aurrekoetxea:2020tuw}, and primordial black hole formation~\cite{deJong:2021bbo, Baumgarte:2026igz, Ning:2026jkk}; 
see Ref.~\cite{Aurrekoetxea:2024ypv} for a recent review.
Other work has also begun to incorporate nonlinear effects into the numerical simulation of cosmological phase transitions~\cite{Ning:2026nfs}.  
In this paper we use \texttt{GABERel}~\cite{Giblin:2019nuv}, which solves Einstein's equations in the \BSSN formulation~\cite{Shibata:1995we,Baumgarte:1998te}, to study a FOPT in the presence of a perfect relativistic radiation fluid~\cite{Baumgarte:2026igz}.  We begin by looking at GUT-scale phase transitions in which bubbles are nucleated near the horizon size, in the presence of this fluid.  

Full numerical relativity requires us to confront the issue of finding proper initial data.  In an FLRW background, it is straightforward to initialize bubbles using the Coleman--de~Luccia bounce~\cite{Coleman:1980aw}, assuming a scalar-field-only universe.  Here, we outline a procedure by which we can satisfy the Hamiltonian and momentum constraints of general relativity on the initial slice by modifying the Coleman--de~Luccia bounce.  We iteratively change the bubble radius to satisfy energy conservation on the initial surface and to match the pre-bounce asymptotic FLRW cosmology.  We will use one- and two-bubble initial states, study the evolution of the energy fractions of the field and fluid, and calculate the resulting gravitational wave spectral density.  Finally, we comment on the existence of oscillons~\cite{Bogolyubsky:1976yu,Gleiser:1993pt,Copeland:1995fq} that appear to form in the wake of these collisions.

The remainder of this paper is organized as follows. In \Sec{sec:model}, we describe our model: a scalar field undergoing a FOPT alongside a radiation fluid, both minimally coupled to gravity. In \Sec{sec:bubsinGABERel}, we describe our numerical implementation, including the construction of constraint-satisfying initial data containing bubbles (\Sec{ssec:initialdata}) and the evolution equations for the metric, field, and fluid (\Sec{ssec:BSSNevolution}). In \Sec{sec:Results}, we present the dynamics of the field and fluid, the resulting gravitational wave spectrum, and the formation of oscillons at late times. We conclude in \Sec{sec:discussion}. Appendices~\ref{appsec:UnitsAndRescaling}, \ref{app:OmegaGW-in-GABERel}, and~\ref{app:dissipation} describe our units and rescaling, the computation of $\Omega_{\rm GW}$ in \texttt{GABERel}, and our choice of numerical dissipation, respectively.

We work in units where $\hbar = c = 1$, but we retain the Planck mass, $\mpl = G^{-1/2} \approx 1.22\times 10^{19}\,{\rm GeV}$. While much of the particle physics literature prefers the reduced Planck mass, $\MPl = \mpl/\sqrt{8\pi}$, we keep explicit factors of $8\pi$ so that our equations are directly comparable to those in the numerical relativity literature, e.g.,~Ref.~\cite{Baumgarte:2010ndz}.
\section{Model}
\label{sec:model}

\begin{figure}
    \centering
    \includegraphics[width=0.99\linewidth]{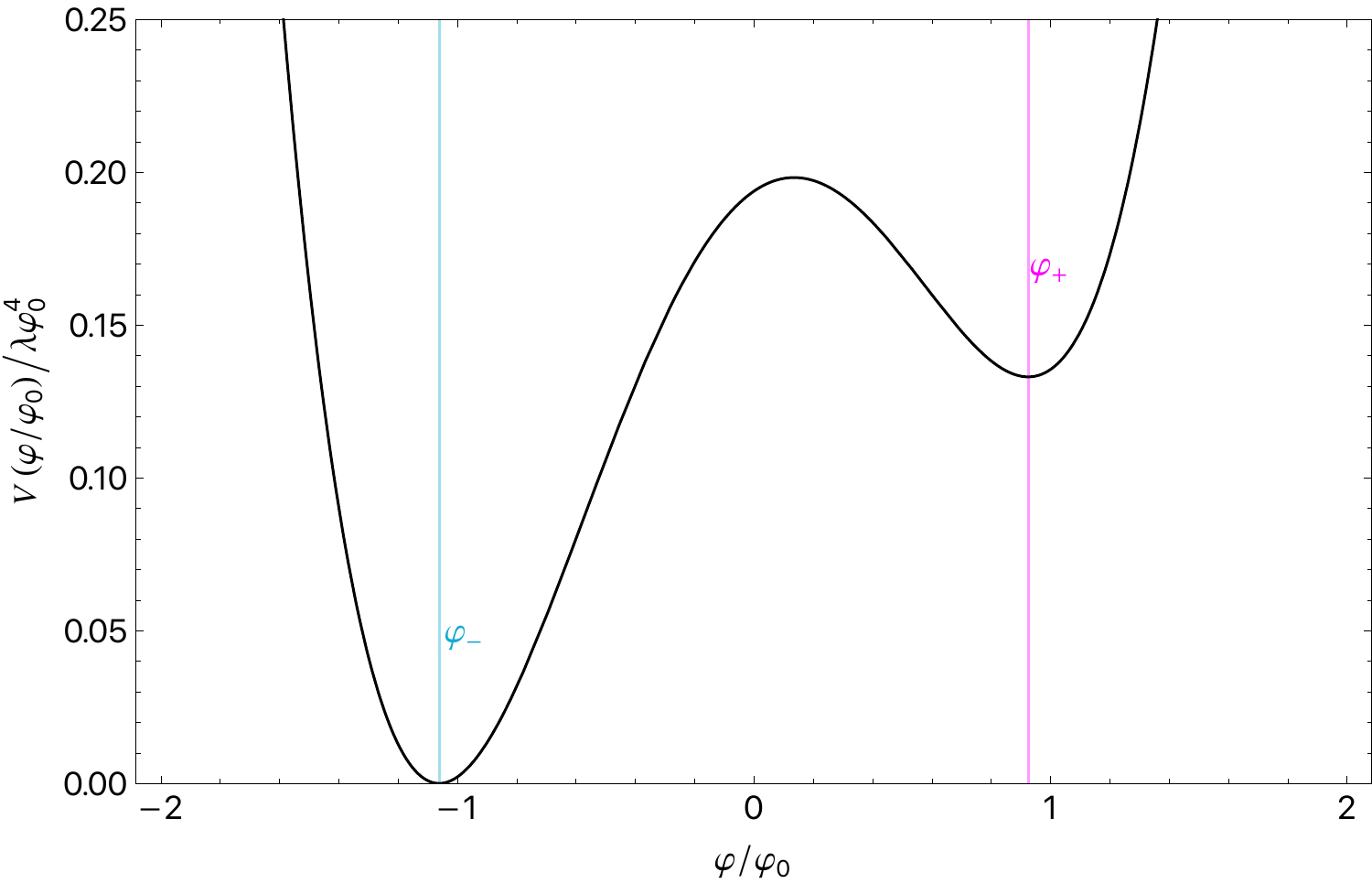}
    \caption{
    \label{fig:potential} The dimensionless potential, $V(\varphi)/\lambda\varphi_0^4$, from \Eq{eq:fullpot}, versus $\varphi/\varphi_0$, where $\epsilon = 1/15$.  The two minima of the potential are located at approximately $\varphi_+ \approx \varphi_0$ (magenta vertical line) and $\varphi_- \approx -\varphi_0$ (cyan vertical line).}
\end{figure}

We consider a universe that has two principal components: a scalar field, $\varphi$, undergoing a FOPT alongside a radiative fluid, with equation of state $p_{\rm fl}/\rho_{{\rm fl},0} = 1/3$.  These components are minimally coupled to gravity as described by the action
\begin{equation}\label{eq:action}
    S = \int {\rm{d}}^4x \sqrt{-g} \left(\frac{\mpl^{2}}{16\pi} R - \frac{1}{2}\partial_\mu \varphi\partial^\mu\varphi - V(\varphi) +\mathcal{L}_{\rm{fluid}} \right),
\end{equation}
where $R$ is the four-dimensional Ricci scalar and $\mathcal{L}_{\rm{fluid}}$ is the Lagrangian of a perfect fluid. 
In this model, 
the field's potential is a tilted double-well potential, as in Ref.~\cite{Child:2012qg}, of the form 
\begin{equation}
\label{eq:fullpot}
    V(\varphi)=\frac{\lambda}{8}(\varphi^2-\varphi_0^2)^2 +\epsilon\lambda\varphi_0^3(\varphi+\varphi_0) + V_0
\end{equation}
where $\lambda$ is the dimensionless self-coupling of the field, 
$\varphi_0$ sets the width of the potential, or rather, the symmetry-breaking scale, 
and $\epsilon$ parametrizes the height difference between the true and false minima.  We additionally include an offset, $V_0$, that ensures that the true minimum of the potential has zero potential energy.  
Once the universe has settled into the true minimum, fluctuations about that minimum have an effective mass, $m^2 \equiv \lambda \varphi^2_0$, which we can also use to set units for our analysis. \Fig{fig:potential} shows a sketch of the potential, \Eq{eq:fullpot}, for a choice of $\epsilon= 1/15$. The field evolves according to the Klein--Gordon equation,
\begin{equation}
    \Box\varphi \equiv \frac{1}{\sqrt{-g}}\partial_\mu\left(\sqrt{-g}\partial^\mu \varphi\right) = \frac{\partial V}{\partial \varphi}.
\end{equation}

For the dynamics of the fluid, we closely follow the notation and dynamics of Ref.~\cite{Baumgarte:2026igz} to consider a perfect radiative fluid defined by its stress-energy,
\begin{equation}
\label{eq:fluidT}
        T^{\mu \nu} = \left(\rho_{{\rm fl},0} + p_{\rm fl} \right) U^{\mu} U^{\nu} + p_{\rm fl} g^{\mu \nu},
\end{equation}
where $\rho_{{\rm fl},0}$ is the rest-frame energy density of the fluid,\footnote{Following Refs.~\cite{Baumgarte:2010ndz,Baumgarte:2026igz}, we use a subscript ``0'' to denote rest-frame quantities, while elsewhere in this paper, e.g., $\Omega_{\rm GW,0}$, a subscript ``0'' denotes quantities evaluated today. We choose to overload our notation in this way to remain consistent with the literature.} $p_{\rm fl}$ is its isotropic pressure, and $U^\mu$ is its four-velocity. We reserve $\rho_{\rm fl} \equiv n_\mu n_\nu T^{\mu\nu}$ for the energy density of the fluid measured by observers normal to our spatial slices (see \Sec{sec:bubsinGABERel}), which differs from $\rho_{{\rm fl},0}$ when the fluid moves relative to those observers.
In this paper, we consider a perfect radiation fluid where $w = p_{\rm fl}/\rho_{{\rm fl},0} = 1/3$.

In our setup, we are interested in phase transitions at a particular energy scale, $\mu$, which corresponds to the energy density of the Universe at the time of the phase transition---the moment when we start our simulations,
\begin{equation}
    \rho_* = \mu^4.
\end{equation}
This energy density is split between the field, which starts with energy density $\rho_{\varphi,*} \approx 2\epsilon\lambda\varphi_0^4$, and the fluid, $\rho_{{\rm fl},*} = \rho_*-\rho_{\varphi,*}$.  Note that the approximate equality of the initial field density is good enough for intuition, but in practice we must calculate the exact $\rho_{\varphi,*}$ from \Eq{eq:fullpot}, which varies by a small but nonzero fraction.  We introduce 
\begin{equation}
    \xi = \frac{2\epsilon \lambda\varphi_0^4 }{ \rho_*}
\end{equation}
that well approximates the ratio of the field's energy density to the total initial energy density at the time of the phase transition.
One can also define the approximate ratio of the field's energy density in the false minimum to the initial energy density of the radiation fluid,
\begin{equation}
    \chi = \frac{2\epsilon \lambda\varphi_0^4}{\rho_{{\rm fl},*}}.
\end{equation}
We use $\chi$ only when comparing with prior work in \Sec{sec:GWs}. 

\section{Computational Strategy\label{sec:bubsinGABERel}}

The simulations presented in the paper use \texttt{GABERel}~\cite{Giblin:2019nuv}, an extension of \texttt{GABE}~\cite{Child:2013ria} that employs fully nonlinear gravity alongside relativistic fields and fluids
in the 
\BSSN formulation~\cite{Shibata:1995we,Baumgarte:1998te}. 

We closely follow the standard textbook treatment~\cite{Baumgarte:2010ndz} of a $3+1$ decomposition.  We start with the Arnowitt--Deser--Misner form of the metric~\cite{Arnowitt:1959ah}
\begin{equation}
    ds^2=-(\alpha^2-\beta^i\beta_i)dt^2+2\gamma_{ij}\beta^i dtdx^j+\gamma_{ij}dx^idx^j
\end{equation}
where $\alpha$ is the lapse, $\beta^i$ is the shift, and $\gamma_{ij}$ is the spatial metric on our three-dimensional spatial hypersurfaces.
To describe the embedding of these hypersurfaces in the four-dimensional manifold, we define the extrinsic curvature 
\begin{equation}
    K_{\mu\nu}=-\gamma^\sigma_\mu \gamma^\kappa_\nu\nabla_\sigma n_\kappa = -\frac{1}{2}\mathcal{L}_n \gamma_{\mu\nu}
\end{equation}
where $\mathcal{L}_n$ is the Lie derivative along the unit normal to the hypersurface
\begin{equation}
    n^\mu = \frac{1}{\alpha}(1,-\beta^i).
\end{equation}
To build our numerical system, we decompose the spatial metric as 
\begin{equation}
    \gamma_{ij}=e^{4\phi}\bar{\gamma}_{ij},
\end{equation}
where $e^{\phi}$ is the conformal factor and $\bar{\gamma}_{ij}$ is the unit-determinant, $\bar{\gamma}=1$, conformal metric.
Similarly, we recast the extrinsic curvature into \BSSN form, 
\begin{equation}
    K_{ij} = e^{4\phi}\tilde{A}_{ij}+\frac{1}{3}\gamma_{ij}K
\end{equation}
where $\tilde{A}_{ij}$ is the trace-free part and $K=\gamma^{ij}K_{ij}$ is the trace of the extrinsic curvature.

Solutions to Einstein's equations must always satisfy the constraint equations
\begin{align}
    \mathcal{H}=& \bar{\gamma}^{ij}\bar{D}_i\bar{D}_je^\phi-\frac{e^\phi}{8}\bar{R}+\frac{e^{5\phi}}{8}\tilde{A}_{ij}\tilde{A}^{ij} \notag\\&-\frac{e^{5\phi}}{12}K^2+\frac{2\pi}{\mpl^2} e^{5\phi}\rho =0 \label{eq:hamconst}\\
    \mathcal{M}^i=& \bar D_j(e^{6\phi}\tilde{A}^{ij})-\frac{2}{3}e^{6\phi}\bar D^i K-\frac{8\pi}{\mpl^2} e^{10\phi}S^i =0 \label{eq:momconst}
\end{align}
where $\mathcal{H}$, \Eq{eq:hamconst}, is the Hamiltonian constraint and $\mathcal{M}^i$, \Eq{eq:momconst}, are the momentum constraints.
Here, $\rho$ and $S^i$ are contractions of the energy-momentum tensor. 
To solve our equations consistently, we must solve the constraint equations on the initial slice.

\subsection{Initial Data}\label{ssec:initialdata}

FOPTs proceed with the nucleation of ``bubbles'' where regions of space instantaneously tunnel from the metastable state to the true minimum.  Extensive work has been done to understand what the field profile would look like in the wake of one of these nucleation events, both without gravity~\cite{Coleman:1977py,Callan:1977pt,Linde:1981zj} and in the presence of gravity~\cite{Coleman:1980aw,Parke:1982pm,Brown:2007sd}.  

In the limit of degenerate minima, the so-called \emph{thin-wall approximation}, each bubble has a profile described by the Coleman--de~Luccia bounce~\cite{Coleman:1980aw} 
\begin{equation}\label{eq:varphi-Coleman}
    \varphi(r,t=0)=\varphi_0\tanh\left(\frac{1}{2}\varphi_0\sqrt{\lambda}\left(r-r_*\right)\right)
\end{equation}
where, in the absence of gravity, the bubble radius can be approximated to be
\begin{equation}
    r_*=(\epsilon\varphi_0\sqrt{\lambda})^{-1}. \label{eq:approx_r}
\end{equation}
In the presence of gravity, Coleman and de Luccia argue that $r_*$ additionally contains a correction.  In the derivation of \Eq{eq:varphi-Coleman}, the authors assume that the scalar field is the only constituent of the universe, so that the false vacuum (de Sitter) and the true vacuum (here Minkowski, since $V=0$ at the true minimum) are both maximally symmetric, time-independent solutions. In the work here, the relativistic fluid breaks this time independence, since the background expands and the fluid dilutes. Its rest frame also picks out a preferred time direction, which breaks the $\mathcal{O}(4)$ symmetry of the Euclidean bounce.  Nonetheless the Coleman--de~Luccia bounce---even in the thin-wall limit---gives us a framework in which to understand the rough relationships between the parameters of the model and the features, e.g., bubble radius and bubble wall thickness, of the nucleation events.

Although we will not use the Coleman--de~Luccia bounce directly as our initial data, we can use these analytic expressions to help inform parameter choice.  From the definition of $\xi$, the self-coupling of the field, $\lambda$, is related to $\epsilon$, $\varphi_0$, and the energy scale $\mu$ via
\begin{equation}
    \lambda = \frac{\xi}{2\epsilon}\left(\frac{\mu}{\varphi_0}\right)^4.
\end{equation}
As a fiducial set of parameters, we start by assuming that $10\%$ of the energy is in the field sector, $\xi = 0.1$, and 
 $\mu = 10^{-4} \mpl$,
corresponding to 
a phase transition at the GUT scale $\sim 10^{13}-10^{15}\,{\rm GeV}$.
We further set $\epsilon = 1/15$ and $\varphi_0 = 2\times 10^{-3} \mpl$, as reasonable choices for the potential.  From a more practical standpoint, these choices allow us to sufficiently resolve the bubble wall with $N^3 = 256^3$ points in a cubic box with each side measuring $L = 1/(2H_*)$.
These parameter choices result in 
a reasonable self-coupling of $\lambda \sim 5\times10^{-6}$. 

Setting up initial bubbles in our scenario requires us to make choices: as discussed above, our setup breaks several of the assumptions of Ref.~\cite{Coleman:1980aw}, so the bounce solution of the Euclidean action is no longer guaranteed to be $\mathcal{O}(4)$-symmetric. Nonetheless, we use this bounce as a starting point for estimating what our bubbles should look like. To set initial data, we employ the following procedure: (1) estimate the hyperspherically symmetric field profile, $\varphi_{\rm B}(r-r_*)$, in flat, four-dimensional Euclidean space, ensuring that the asymptotic field is in the false vacuum; (2) use flat-space conservation of energy to obtain an initial estimate of the bubble radius, $r_*$; and (3) solve the Hamiltonian and momentum constraints, \Eqs{eq:hamconst}{eq:momconst}, for the gravitational degrees of freedom, iteratively adjusting $r_*$ as described below. In practice, the momentum constraints are trivially satisfied, since the field and fluid both begin at rest and we choose $\tilde{A}_{ij} = 0$ and a spatially uniform $K$ on the initial slice.

As a first guess, we assume that the shape of the bubble wall is set mainly by the dynamics of the scalar field. We therefore follow Refs.~\cite{Coleman:1977py,Callan:1977pt, Coleman:1980aw}, temporarily assuming flat spacetime and a universe containing only the scalar field. In this case we look for $\mathcal{O}(4)$-symmetric field configurations $\varphi(s)$, where $s$ is the four-dimensional Euclidean radius, which satisfy
\begin{equation}
    \frac{d^2\varphi}{ds^2}+\frac{3}{s}\frac{d\varphi}{ds}=\frac{dV}{d\varphi}, \label{eq:field_motion}
\end{equation}
subject to regularity at the origin, $d\varphi/ds|_{s=0}=0$, and $\varphi\rightarrow\varphi_+$ (the false vacuum) as $s\rightarrow\infty$. We solve \Eq{eq:field_motion} with a shooting method, tuning the central value $\varphi(0)$, which lies close to the true minimum, between solutions that overshoot and undershoot the false vacuum. This method reliably determines the shape of the wall. However, near the thin-wall limit the location of the wall, $s_*$, is extremely sensitive to $\varphi(0)$. In any case, the flat-space value need not be the correct radius once gravity and the fluid are included. We therefore keep only the wall profile and fix the radius separately, as described below. On the $t=0$ slice, $s = r$, and time-reflection symmetry guarantees $\partial_t\varphi=0$, so we take $\varphi_{\rm B}(r-r_*)$, with the field at rest, as the initial profile of our bubbles.

Next, we require that bubble nucleation conserve the total (flat-space) energy of the scalar field. That is, we impose
\begin{equation}
    \Delta E(r_*)=\int d^3x\left[\frac{(\nabla\varphi_{\rm B})^2}{2}+V\left(\varphi_{\rm B}\right)-V_+\right]=0,
\end{equation}
where $V_+$ is the value of the potential in the false minimum. For a fixed wall profile, $\Delta E$ depends only on the bubble radius $r_*$. In the thin-wall limit, this condition reproduces \Eq{eq:approx_r} exactly, since the energy released in the bubble interior, $4\pi r_*^3\,\Delta V/3$, where $\Delta V = V_+ - V_- \approx 2\epsilon\lambda\varphi_0^4$, balances the energy in the wall, $4\pi r_*^2\sigma$, where $\sigma = \tfrac{2}{3}\sqrt{\lambda}\,\varphi_0^3$ is the wall tension. For our finite-thickness walls, we start from \Eq{eq:approx_r} and adjust $r_*$ iteratively until $\Delta E = 0$.

For simulations with more than one bubble, we place bubbles on the grid by randomly generating $n$ bubble centers on a grid with $N^3$ points. To ensure that bubble walls do not overlap, each new center must lie a distance $d > 2.5\,r_*$ from those previously placed (measured across the periodic boundaries), where $r_*$ is the thin-wall estimate, \Eq{eq:approx_r}.

Finally, we must make the bubbles consistent with Einstein's equations by determining the geometry of the initial slice from the constraints, \Eqs{eq:hamconst}{eq:momconst}. We assume the field, as determined above, is instantaneously at rest at the time of nucleation and that the radiation fluid is homogeneous. We choose the conformal metric to be flat, $\bar{\gamma}_{ij} = \delta_{ij}$, and the extrinsic curvature to be pure trace and spatially homogeneous,
\begin{equation}
K_{ij}=\frac{1}{3}\gamma_{ij}K,
\end{equation}
with $\bar{D}^iK=0$. With $\tilde{A}_{ij}=0$, constant $K$, and vanishing momentum density, the $\mathcal{M}^i$ are identically zero, and the Hamiltonian constraint, \Eq{eq:hamconst}, reduces to
\begin{equation}
\nabla^2\psi=-\frac{2\pi}{\mpl^2}\psi^5\rho+\frac{\psi^5}{12}K^2,
\label{eq:hamflat}
\end{equation}
where $\psi\equiv e^{\phi}$ is the conformal factor and $\rho$ includes both the field and the fluid. 

For this equation to be satisfied in a periodic domain, the divergence theorem requires that the Laplacian has zero mean, $\langle \nabla^2\psi\rangle=0$.  Averaging \Eq{eq:hamflat} over the box shows that a solution exists only if
\begin{equation}
K=-\sqrt{\frac{24\pi}{\mpl^2}\frac{\langle\rho\psi^5\rangle}{\langle\psi^5\rangle}}\label{eq:meanK}.
\end{equation}
This fixes $K$ in terms of a $\psi^5$-weighted mean of the energy density.  This fact complicates our picture of the bounce solution. The instant before nucleation, the universe is homogeneous and the homogeneous value of the expansion is given by the FLRW solution to Einstein's equations,
\begin{equation}
\frac{K^2}{9}=H_*^2=\frac{8\pi\rho_*}{3\mpl^2}\label{eq:Friedmann}.
\end{equation}

Nucleation is a local event, and regions of space outside its future light cone cannot respond to it. We therefore expect the expansion rate far from the bubble to remain the pre-nucleation value, \Eq{eq:Friedmann}. On our constant-$K$ slice, this requires
\begin{equation}
\frac{\langle\rho\psi^5\rangle}{\langle\psi^5\rangle} = \rho_*.
\label{eq:matching}
\end{equation}
Our flat-space construction, however, conserves the \emph{unweighted} mean, $\langle\rho\rangle = \rho_*$. Because the bubble curves the space it occupies, $\psi \neq 1$, and the two averages differ, so the flat-space bubble cannot satisfy both the Hamiltonian constraint and \Eq{eq:matching}.

The mismatch arises because both the shape and the radius of the bubble were computed in flat space. In reality, gravity modifies both.  Here we assume that the dominant correction is to the radius, as in the thin-wall limit of the scalar-only system studied by Coleman and de~Luccia~\cite{Coleman:1980aw}, where gravity shifts the bubble radius but leaves the wall profile unchanged. 
Starting from the flat-space radius found above, we therefore adjust $r_*$ iteratively until the Hamiltonian constraint and the matching condition, \Eq{eq:matching}, are both satisfied within a set tolerance.

The constant mode of $\psi$ needs separate treatment. For homogeneous $\rho$, any constant $\psi$ solves \Eq{eq:hamflat}, reflecting the freedom to choose the scale factor on the initial slice. This degeneracy persists for inhomogeneous $\rho$ when $K$ is determined by \Eq{eq:meanK}. We fix it by requiring $\langle\psi\rangle=1$ and solve
\begin{equation}
\mathcal{G}=-\nabla^2\psi-\frac{2\pi}{\mpl^2}\psi^5\rho+\frac{\psi^5}{12}K^2+\langle\psi\rangle-1=0.\label{eq:Gcode}
\end{equation}
Here $K$ is recomputed from \Eq{eq:meanK} at each step, so that the mean of $\mathcal{G}$ reduces to $\langle\psi\rangle - 1$.
We compute the analytic Jacobian of $\mathcal{G}$ using the spectral Laplacian and solve for $\psi$ using Newton's method with this exact Jacobian. Comparing the output value of $K$ to \Eq{eq:Friedmann} then allows us to determine the direction to perturb $r_*$, and we alternate solving \Eq{eq:Gcode} with making small corrections to the radius of the bubble. Upon completion, the initialization routine returns the configuration of the field and conformal factor on the lattice and the corresponding homogeneous value of $K$. 

\begin{figure}[t]
    \centering
    \includegraphics[width=0.99\linewidth]{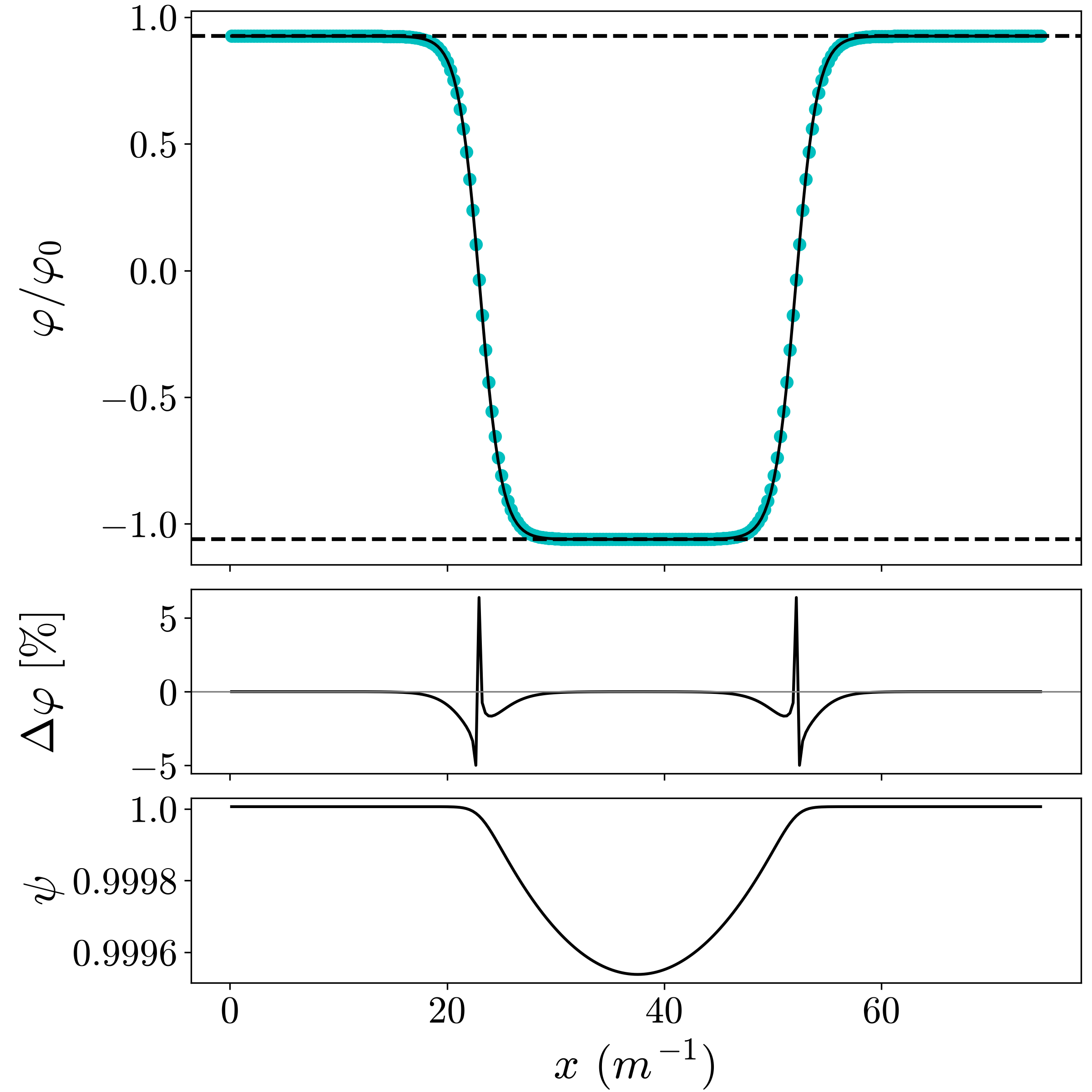}
    \caption{Initial spatial profiles 
    generated from our initial condition routine. 
    (Top) Comparison of our initial field profile (cyan circles) to the Coleman--de~Luccia bounce solution (solid black) normalized by $\varphi_0$.
    We indicate the field values at the true and false minima, $\varphi_-$ and $\varphi_+$, with the dashed black lines. 
    (Middle) Percent difference between the numerical and analytical bubble profiles. 
    (Bottom) Conformal factor profile computed from solving the Hamiltonian constraint. 
    }
    \label{fig:initial_profile}
\end{figure}

While our initial data solver does not initialize the bubble using the approximate thin-wall solution in \Eq{eq:varphi-Coleman}, it is worthwhile to compare our solutions to this thin-wall solution.
The top panel of \Fig{fig:initial_profile} shows this comparison where the cyan circles are our numerical initial data and the solid black line is the approximate solution described by \Eq{eq:varphi-Coleman}.
To compare how the shape of the profiles differs, we shift and rescale the analytical profile to be aligned with the field values at the true and false minima $\varphi_-$, $\varphi_+$ rather than $-\varphi_0$ and $+\varphi_0$.
In the middle panel, we present a percent difference between the two profiles, $\Delta \varphi = 100\times(\varphi_{\rm num}-\varphi_{\rm ana})/\varphi_{\rm ana}$ to better highlight where the profiles differ. 

Inside and outside of the bubble walls, the profiles match within a fraction of a percent or less. 
At the bubble walls, there is a larger difference.
We note, however, that since the bubble profile passes through zero, we see a spike in the middle panel due to division by zero. Away from this point, we find that the profiles differ by $\lesssim 2\%$ in the walls, indicating good agreement between our numerical profile and the thin-wall approximate solution.
Finally, in the bottom panel of \Fig{fig:initial_profile}, we present the profile for the conformal factor, $\psi$, found from solving the Hamiltonian constraint. 
Outside the bubble walls, the conformal factor is just above $1$,
and inside the bubble, the conformal factor decreases from $1$ by $\approx 0.04\%$. 

\subsection{Evolution Equations}\label{ssec:BSSNevolution}
In addition to the constraint equations, the $3+1$ decomposition of Einstein's equations results in a set of evolution equations for the metric quantities,
\begin{align}
    \partial_t \phi &= -\frac{1}{6}\alpha K + \beta^i \partial_i \phi + \frac{1}{6}\partial_i \beta^i \\
    \partial_t \bar{\gamma}_{ij} &= -2\alpha \tilde{A}_{ij} + \beta^k \partial_k \bar{\gamma}_{ij} + \bar{\gamma}_{ik}\partial_j \beta^k \notag \\
    &\quad + \bar{\gamma}_{kj}\partial_i \beta^k - \frac{2}{3}\bar{\gamma}_{ij}\partial_k \beta^k \\
    \partial_t K &= -\gamma^{ij} D_j D_i \alpha + \alpha \left( \tilde{A}_{ij}\tilde{A}^{ij} + \frac{1}{3}K^2 \right) \notag \\
    &\quad + \frac{4\pi}{\mpl^2}\alpha (\rho + S) + \beta^i \partial_i K \\
    \partial_t \tilde{A}_{ij} &= e^{-4\phi}\left(-D_j D_i \alpha + \alpha \left(R_{ij} - \frac{8\pi}{\mpl^2} S_{ij}\right)\right)^{TF} \notag \\
    &\quad + \alpha \left( K\tilde{A}_{ij} - 2\tilde{A}_{il}\tilde{A}^l_j \right) + \beta^k \partial_k \tilde{A}_{ij} \notag \\
    &\quad + \tilde{A}_{ik}\partial_j \beta^k + \tilde{A}_{kj}\partial_i \beta^k - \frac{2}{3}\tilde{A}_{ij}\partial_k \beta^k,
\end{align}
where $\rho = n^\mu n^\nu T_{\mu \nu}$, $S_i = -\gamma_{i\mu}n_\nu T^{\mu \nu}$ (used in \Eq{eq:momconst}), $S_{ij} = \gamma_{i\mu}\gamma_{j\nu}T^{\mu \nu}$, and $S=\gamma^{ij}S_{ij}$ are contractions of the energy-momentum tensor.
Namely, the energy density is given as
\begin{align}
    \rho &= n^\mu n^\nu T_{\mu\nu}\nonumber \\
        &= \frac{1}{2}\left(\Pi^2+\nabla_i\varphi\nabla^i\varphi\right)+V(\varphi)+\rho_{\rm fl} 
\end{align}
where $\rho_{\rm fl} = \gamma^{-1/2}E$; $\Pi$ and $E$ are defined below.

To evolve the scalar field, we introduce an auxiliary variable $\Pi = \mathcal{L}_n\varphi$ to describe the momentum of the field.
Together, the field and its momentum evolve as
\begin{align}
    \Pi &= \frac{1}{\alpha}\left( \partial_t \varphi - \beta^i \partial_i \varphi \right) \\
    \partial_t \Pi &= \beta^i \partial_i \Pi + \gamma^{ij}\partial_i\alpha\,\partial_j\varphi \notag \\
    &\quad + \alpha\left( \gamma^{ij}\partial_i\partial_j\varphi - \gamma^{ij}\Gamma^k_{ij}\partial_k\varphi + K\Pi - \frac{\partial V}{\partial\varphi}\right).
\end{align}
We note that we choose a sign convention for $\Pi$ that is consistent with Ref.~\cite{Giblin:2019nuv} but opposite to that of Ref.~\cite{Baumgarte:2010ndz}.

To model the perfect fluid, we use the scheme described in Ref.~\cite{Baumgarte:2026igz}.
Under conservation of stress-energy ($\nabla_\mu T^{\mu\nu} = 0$) and given \Eq{eq:fluidT}, we find that the fluid's coordinate energy density, $E \equiv \sqrt{\gamma}n_\mu n_\nu T^{\mu\nu}$, and coordinate momentum, $P^i \equiv -\sqrt{\gamma}\,n_\nu T^{i\nu}$, evolve as 
\begin{align}
\partial_t E & + \alpha \left(\partial_i P^i\right) - \beta^i(\partial_i E) - E\left(\partial_i \beta^i\right) \label{eq:Edensity}\\
&= -2P^i(\partial_i \alpha) + P^i \beta^j K_{ij} + K_{ij} S_{\rm v}^{ij} - P^j \beta^i K_{ij}, \nonumber
\end{align}
and
\begin{align}
\partial_t P_j & + \gamma_{kj}\, \partial_i S_{\rm v}^{ik} + S_{\rm v}^{ik}(\partial_i \gamma_{kj}) - P_j(\partial_i \beta^i) - \beta^i(\partial_i P_j) \nonumber \\
&= -E\left(\partial_j \alpha\right) + \frac{P^i}{2}\beta^k(\partial_j \gamma_{ik}) + P_k(\partial_j \beta^k) \nonumber\\
&\quad + \frac{1}{2}S_{\rm v}^{ik}(\partial_j \gamma_{ik}) - \frac{1}{2}P^k \beta^i(\partial_j \gamma_{ik})\label{eq:Momdensity}. 
\end{align}
where $S_{\rm v}^{ij} \equiv \alpha\sqrt{\gamma}\,S^{ij}_{\rm fl}$ is the densitized fluid stress (see Eq.~(A18) of Ref.~\cite{Baumgarte:2026igz}).

Following Ref.~\cite{Baumgarte:2026igz}, we derive these equations in the spirit of the Wilson scheme~\cite{Wilson:1972,Hawley:1984}. They are closely related to the flux-conservative ``Valencia'' formulation used in high-resolution shock-capturing schemes~\cite{Banyuls:1997zz,Font:1998hf} (see also Ref.~\cite{Baumgarte:2010ndz}). That formulation evolves the conserved rest-mass density $\tilde D = \gamma^{1/2}W\rho_{\rm m}$, where $W$ is the Lorentz factor of the fluid relative to normal observers, momentum density $\tilde S_j = \alpha\gamma^{1/2}{T^0}_j$, and energy density $\tilde\tau = \alpha^2\gamma^{1/2}T^{00}-\tilde D$, where $\rho_{\rm m}$ is the rest-mass density. For a radiation fluid there is no conserved particle number, so the equation for $\tilde D$ is not needed. Our evolved variables then correspond to $E = \tilde\tau + \tilde D$ and $P_j = \tilde S_j$, and \Eqs{eq:Edensity}{eq:Momdensity} are equivalent to the Valencia energy and momentum equations, with our $\rho_{{\rm fl},0}$ playing the role of the total rest-frame energy density, $\rho_{\rm m}(1+e_{\rm int})$, where $e_{\rm int}$ is the specific internal energy. As a result, spatial derivatives act only on evolved variables (and combinations of them), and the primitive variables $\rho_{{\rm fl},0}$ and $v^i$, the fluid three-velocity, are recovered algebraically on each slice. Unlike Ref.~\cite{Font:1998hf}, we do not use a Riemann solver, but evolve all metric, field, and fluid variables with the same finite-difference scheme. This is adequate here because the fluid couples to the scalar field only through gravity, and we see no evidence of shock formation in our simulations.

Our lapse evolves according to a Bona-Mass\'o slicing condition \cite{Bona:1994dr}, a modified $1+\log$ slicing condition, adopted in Ref.~\cite{Giblin:2019nuv},
\begin{equation}
    (\partial_t -\beta^i \partial_i)\alpha = -2\alpha(K-\left<K\right>),
\end{equation}
and the shift evolves following the Gamma-driver condition
\begin{align}
\partial_t \beta^i &= \frac{3}{4} B^i \\
\partial_t B^i &= \partial_t \bar{\Gamma}^i - \eta B^i
\end{align}
with $\eta = 100\,m$ and 
\begin{align}
    \partial_t \bar{\Gamma}^i &= -2\tilde{A}^{ij}\partial_j \alpha + 2\alpha\left( \bar{\Gamma}^i_{jk}\tilde{A}^{kj} - \frac{2}{3}\bar{\gamma}^{ij}\partial_j K \right. \notag \\
    &\quad \left. - \frac{8\pi}{\mpl^2} \bar{\gamma}^{ij}S_j + 6\tilde{A}^{ij}\partial_j \phi \right) + \beta^j \partial_j \bar{\Gamma}^i \notag \\
    &\quad - \bar{\Gamma}^j \partial_j \beta^i + \frac{2}{3}\bar{\Gamma}^i \partial_j \beta^j + \frac{1}{3}\bar{\gamma}^{li}\partial_l \partial_j \beta^j \notag \\
    &\quad + \bar{\gamma}^{lj}\partial_j \partial_l \beta^i.
\end{align}

To evolve our simulations on a lattice,
we use fourth-order finite-difference stencils and standard fourth-order Runge--Kutta time integration with a Courant factor $\Delta t/\Delta x = 1/30$. 
To remove high-frequency noise, we employ Kreiss--Oliger dissipation~\cite{KreissOliger1973} and set the dissipation parameter to $\nu_{\rm dis}=0.4$.
We find that this value sufficiently dissipates high-frequency noise while leaving the physics unaffected; for more details, see \App{app:dissipation}. 

As a final note, we rescale all of our dimensionful quantities to be dimensionless in our simulations. 
Throughout this paper, we express times and lengths in units of $m^{-1}$, where $m = \sqrt{\lambda}\varphi_0$ is the effective mass defined in \Sec{sec:model}. For more details, see \App{appsec:UnitsAndRescaling}.

\begin{figure*}[t!]
    \centering
    \includegraphics[width=0.245\textwidth]{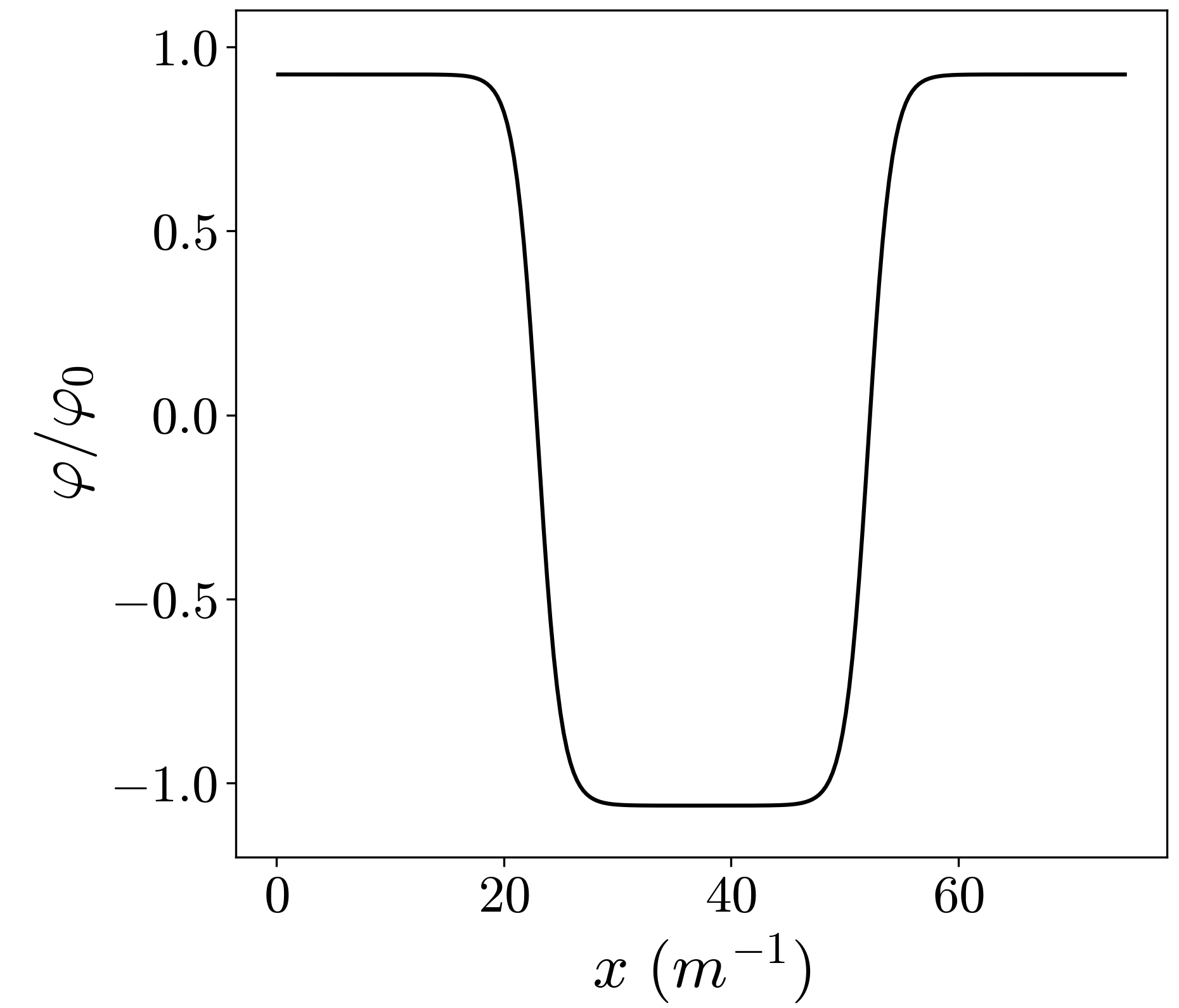}
    \includegraphics[width=0.245\textwidth]{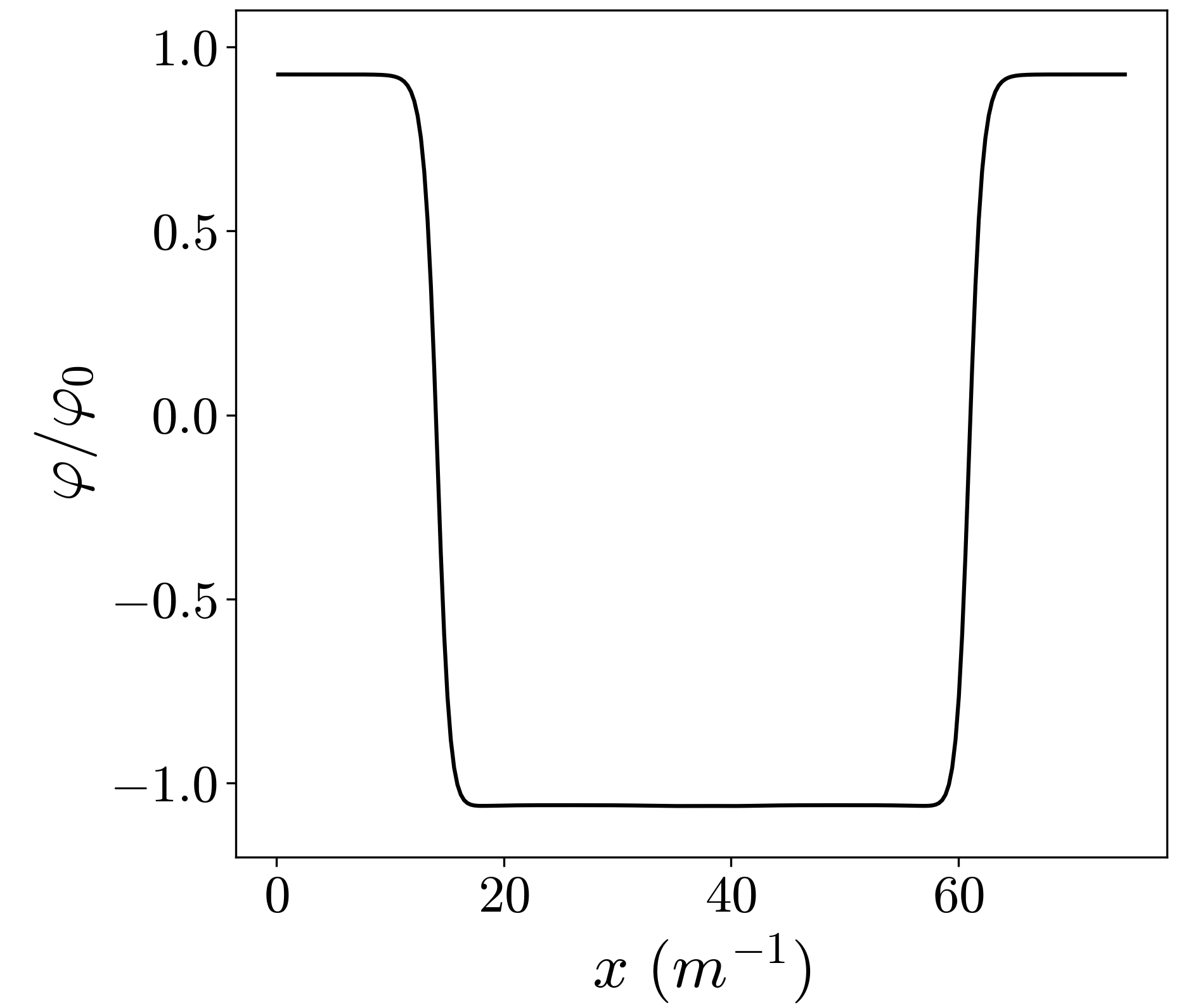}
    \includegraphics[width=0.245\textwidth]{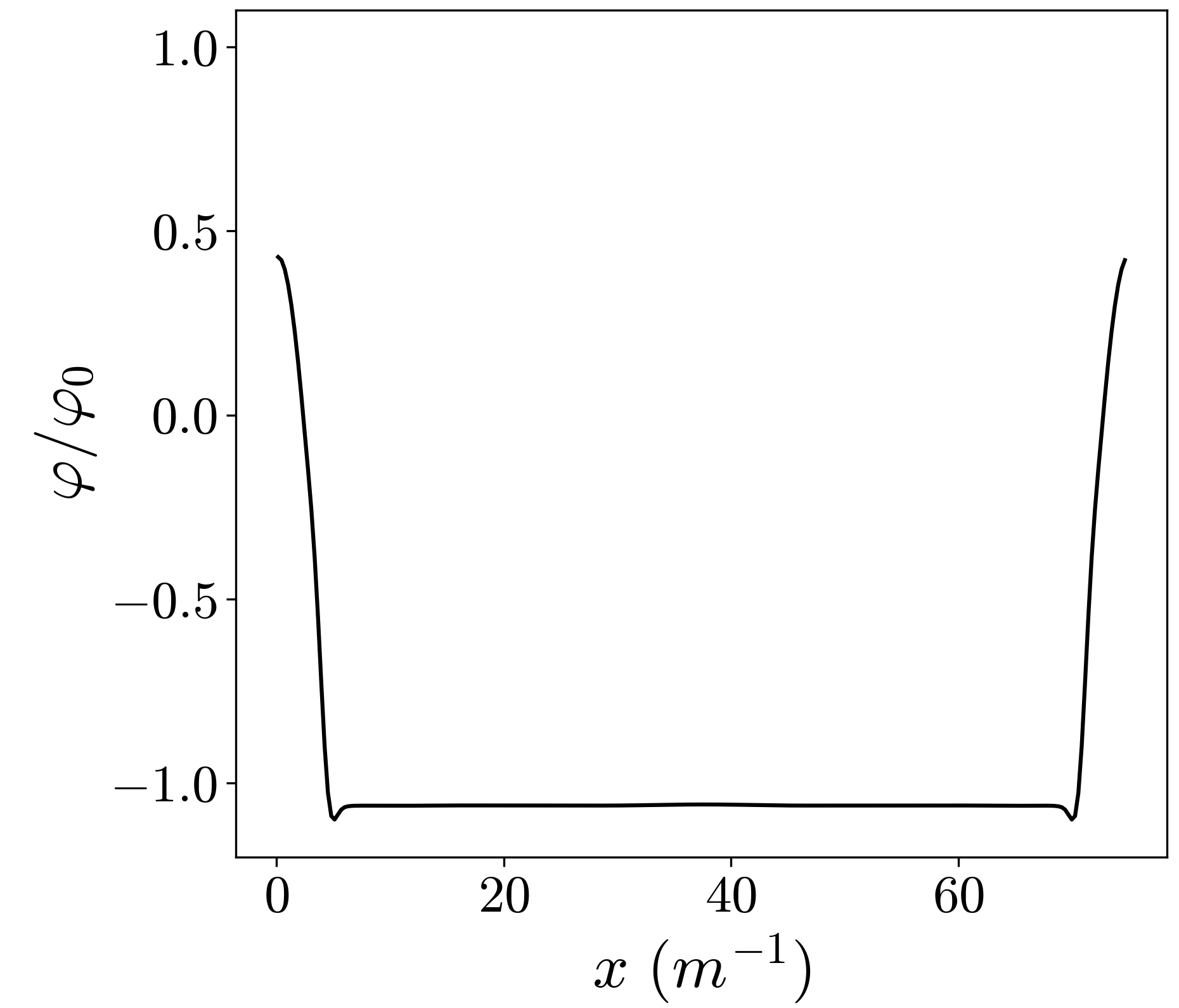}
    \includegraphics[width=0.245\textwidth]{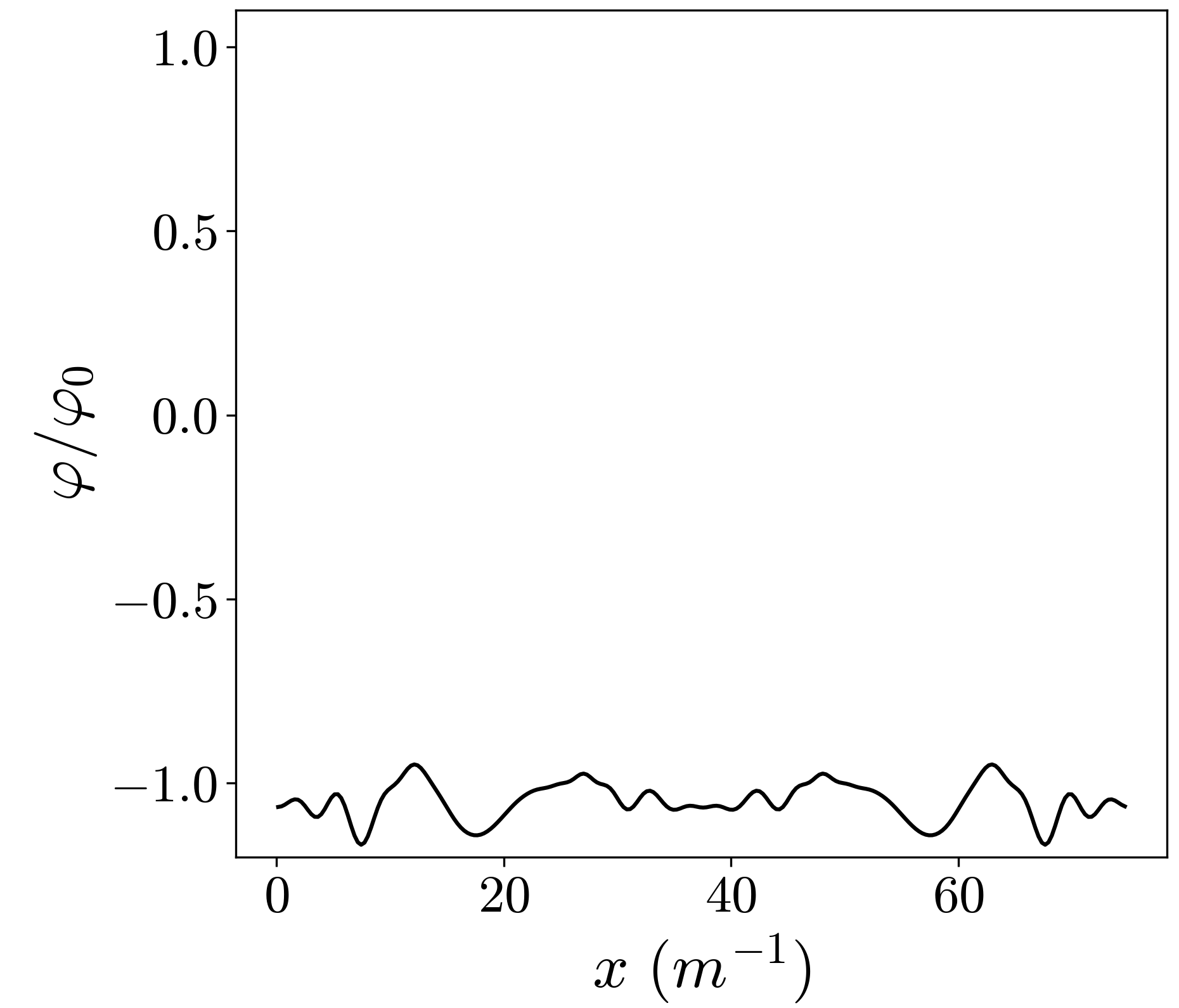}\\
    \includegraphics[width=0.245\textwidth]{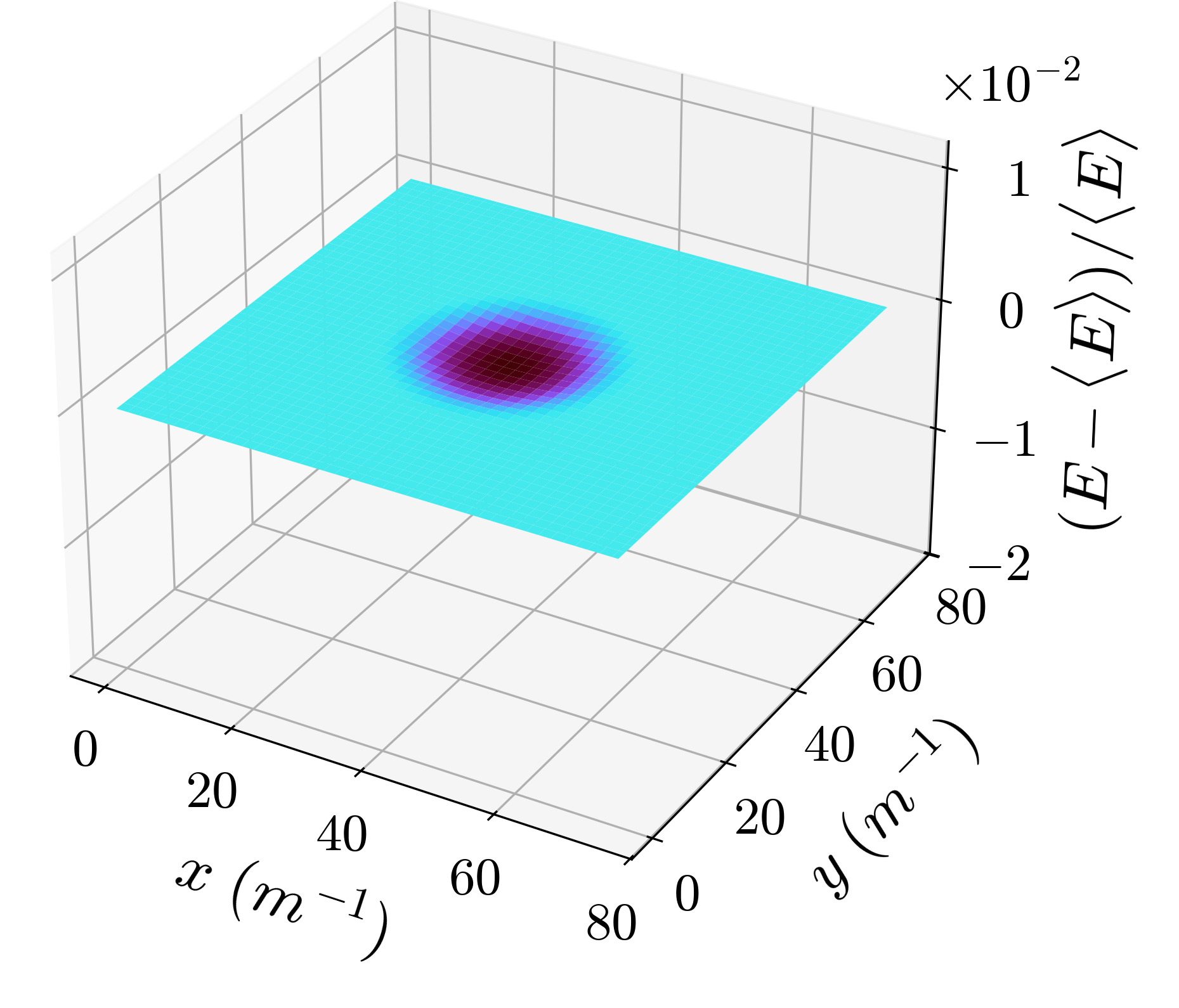}
    \includegraphics[width=0.245\textwidth]{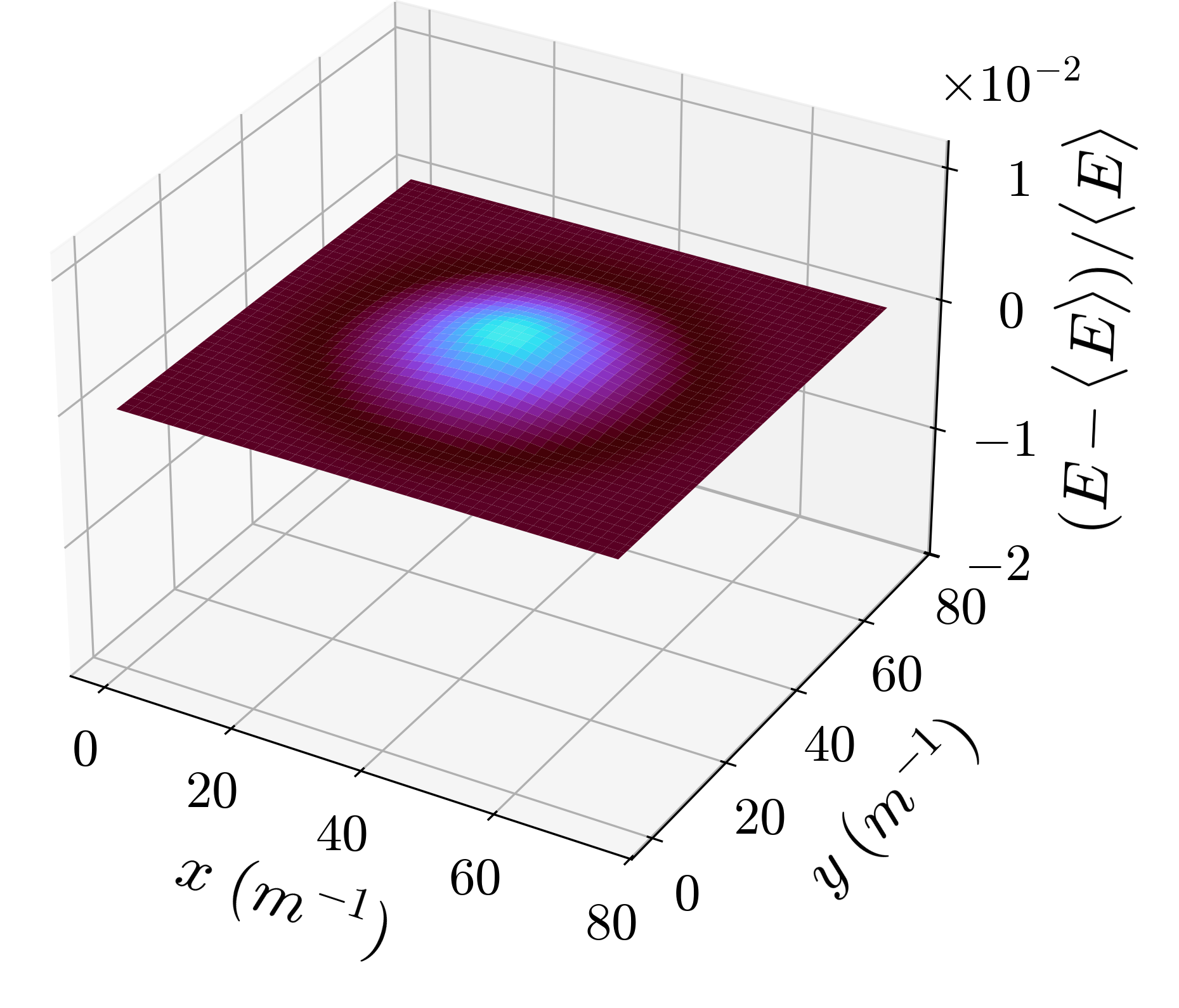}
    \includegraphics[width=0.245\textwidth]{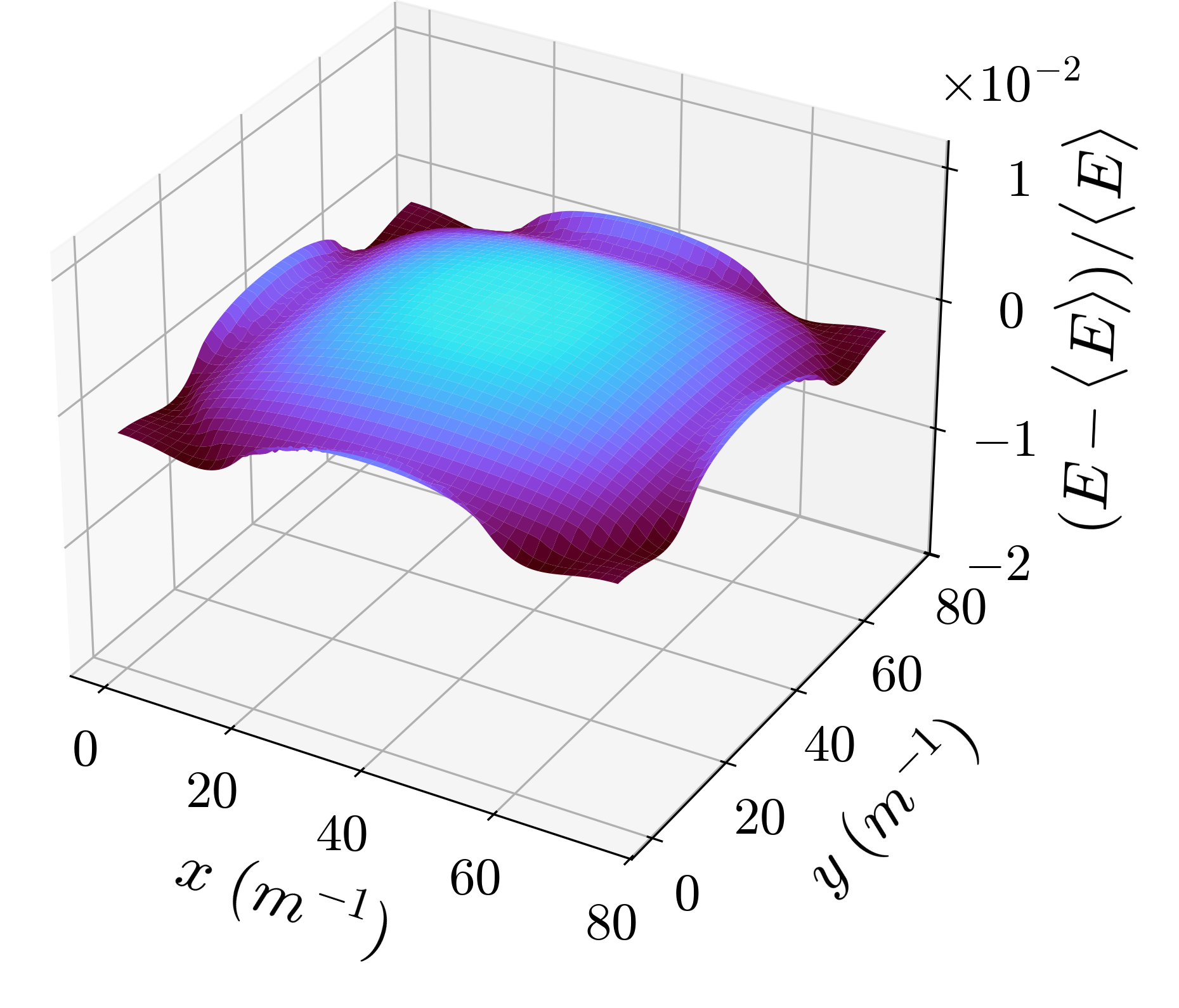}
    \includegraphics[width=0.245\textwidth]{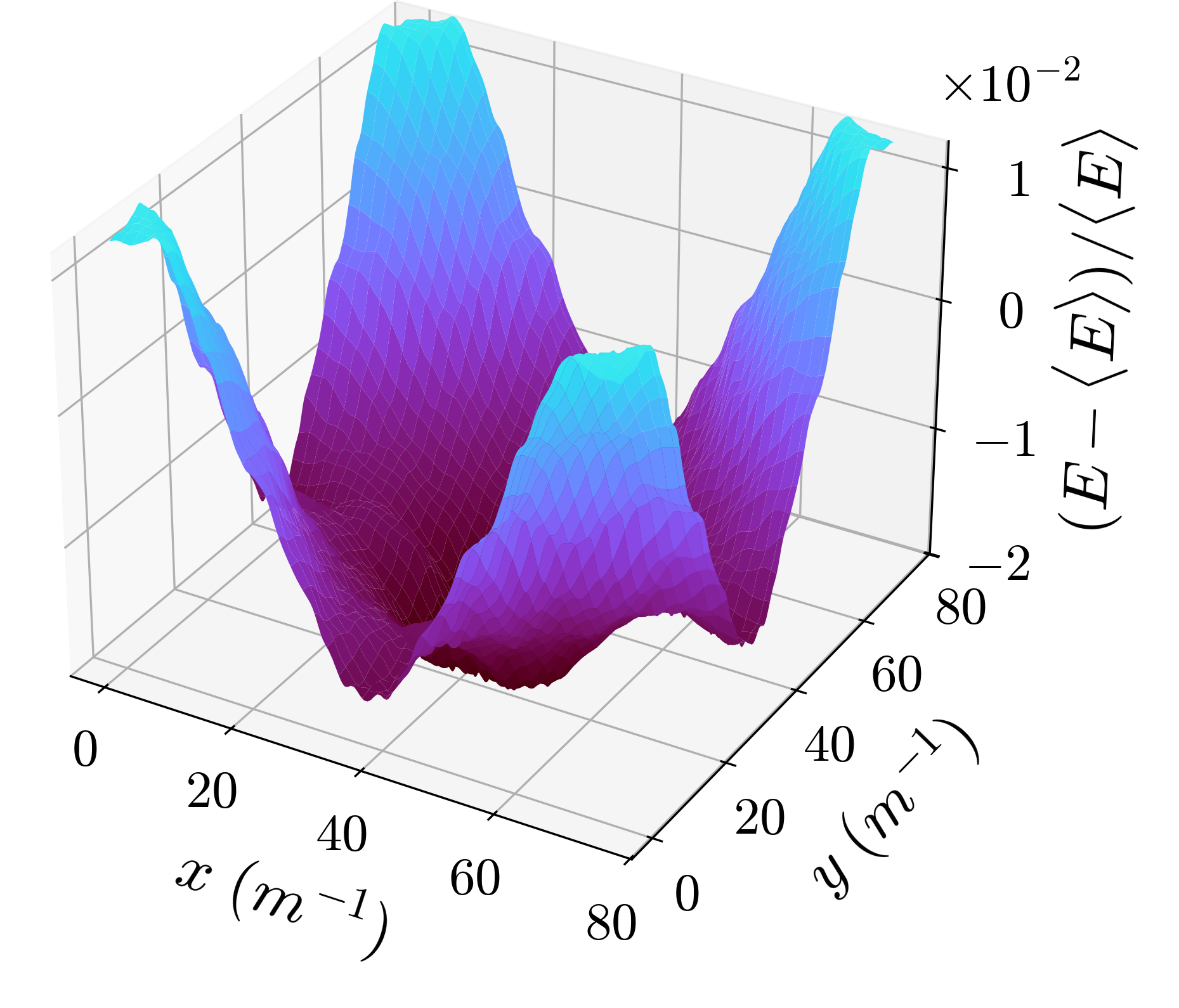}
    \caption{Snapshots of the field and fluid evolution at $t = 0$ (A), $t= 19.47\,m^{-1}$ (B), $t=43.82\,m^{-1}$ (C), and $t=116.9\,m^{-1}$ (D) (from left to right) where A B, C, and D correspond to labels in Fig.~\ref{fig:rho_vs_a}.
    (Top) One-dimensional slices of the normalized field $\varphi/\varphi_0$ through the center of the grid along the $x$-axis.
    (Bottom) The relative change of the fluid energy density, $(E-\langle E\rangle)/\langle E \rangle$, on a two-dimensional slice through the center of the grid, shown as a surface.
    }
    \label{fig:phi_E_panels}
\end{figure*}
\section{Results}\label{sec:Results}

We now turn to the evolution of our fiducial model.
As a brief summary of our results, we find that as the bubble expands and the field eventually settles into its true minimum, the fluid energy density decreases over time and the gravitational wave power spectrum grows due to the collision of the bubble walls.

\subsection{Bubble and Fluid Dynamics}

We begin by focusing on the scenario where there is just one bubble per box; since the box size is set to $L = 1/(2H_*)$, this corresponds to about 8 bubbles per Hubble volume, $H_*^{-3}$.  Since our box is periodic, we center the bubble in the middle of the grid without any loss of generality.  As described in \Sec{ssec:initialdata}, the field is initialized at rest. The fluid is initialized at rest with a homogeneous density, $\rho_{\rm fl} = \rho_{{\rm fl},*}$; however, the coordinate energy density, $E = e^{6\phi}\rho_{\rm fl}$, is not homogeneous.

Figure~\ref{fig:phi_E_panels} shows a one-dimensional slice of the field $\varphi/\varphi_0$ (top) and a two-dimensional slice of the relative change of the fluid energy density $(E-\langle E\rangle)/\langle E \rangle$ (bottom) corresponding to the initial time (leftmost panels), an early time before collisions begin (second panels), the beginning of wall collisions (third panels), and once the field has settled into its true minimum (rightmost panels).  We will reference these four times as (A), (B), (C), and (D) respectively.

Initially (A, leftmost panels), the bubble, as shown in \Fig{fig:initial_profile}, has a radius of approximately $r_* \approx 15\,m^{-1} \approx 0.2L$. 
In the region exterior to the bubble, the fluid's coordinate energy density corresponds to the average energy density of the simulation, since $\phi \approx 0$ in this region. Along the bubble wall and inside the bubble, the fluid's coordinate energy density $E$ is slightly lower, even though $\rho_{\rm fl}$ is constant across the slice.
After the bubble has evolved for some time, at $t = 19.47\,m^{-1}$ (B, second panels), we can see that the bubble walls have become thinner; that is, the gradient of the field across the walls has steepened.
As the bubble expands, more of the volume settles into the true minimum, and the released vacuum energy is transferred into the bubble walls.
Consequently, we see larger fluctuations in the fluid coordinate energy density, $E$.

By $t = 43.82\,m^{-1}$ (C, third panels), the bubble walls have started to collide. 
As this occurs we see larger fractional variations in the fluid. 
Finally, the bubble walls cease colliding around $t = 116.9\,m^{-1}$ (D, rightmost panels), and the field oscillates about its true minimum, indicating the phase transition is completing. 
By this time, the fluid is beginning to show percent-level variations in $E$---indicating that the fluid is being churned in response to the inhomogeneous metric.

While the relative change of the fluid's energy density $E$ increases as shown in the bottom panels of \Fig{fig:phi_E_panels}, the overall magnitude of $E$ decreases throughout the simulation---this is due to the dilution of the fluid as the volume of the simulation expands.  At the beginning of the simulation the relative contribution of the fluid, $\Omega_{\rm fl} = \rho_{\rm fl}/\rho$, is set to approximately ninety percent, $\Omega_{{\rm fl},*} \sim 1-\xi \approx 0.9$, and we expect this to decline somewhat due to the fact that $\rho_{\rm fl}$ dilutes as $a^{-4}$.  The field, on the other hand, can contribute as vacuum energy (in the regions where the field is still in the false minimum) or as matter or radiation depending on the wavelength of the specific modes that are excited as the field falls into its true minimum.  

\begin{figure}
    \centering
    \includegraphics[width=0.99\linewidth]{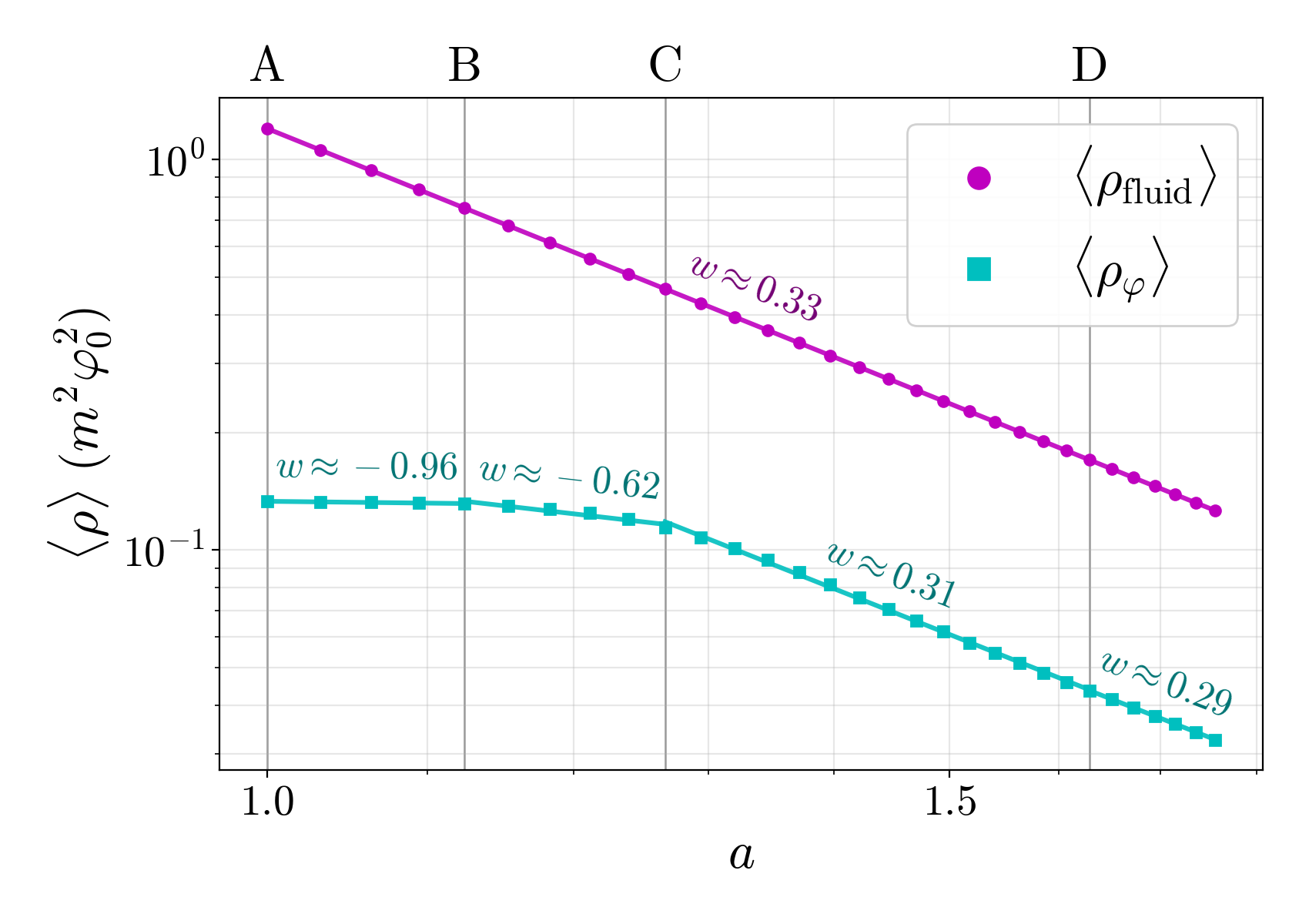}
    \caption{Evolution of the volume-averaged energy densities of the fluid, $\langle\rho_{\rm fl}\rangle$ (magenta dots), and the field, $\langle\rho_{\varphi}\rangle$ (cyan squares), as a function of the effective scale factor $a$ (normalized to unity at $t=0$).
    The vertical lines labeled A, B, C, and D correspond to $t = 0$, $t= 19.47\,m^{-1}$, $t=43.82\,m^{-1}$, and $t=116.9\,m^{-1}$ (from left to right), the same times as in \Fig{fig:phi_E_panels}. 
    We provide lines of best fit for each time interval and compute the effective equation of state parameter $w$ by fitting $\rho \propto a^{-3-3w}$. 
    }
    \label{fig:rho_vs_a}
\end{figure}

We present the evolution of the volume-averaged energy densities of the field and fluid, $\langle\rho_\varphi\rangle$ and $\langle\rho_{\rm fl}\rangle$, where for any quantity $X$
\begin{equation}
    \langle X\rangle = \frac{\int d^3x\sqrt{\gamma}\, X(x)}{\int d^3x\sqrt{\gamma}},
\end{equation}
versus the \emph{effective} scale factor
\begin{equation}
    a = \left(\frac{\int d^3x\sqrt{\gamma}}{\int d^3x\sqrt{\gamma}\,\big|_{t=0}}\right)^{1/3},
\end{equation}
in \Fig{fig:rho_vs_a}. 
Vertical lines labeled A, B, C, and D correspond to the same times as in \Fig{fig:phi_E_panels}: $t = 0$, $t= 19.47\,m^{-1}$, $t=43.82\,m^{-1}$, and $t=116.9\,m^{-1}$ (from left to right).
We estimate the equation of state parameter $w$ from fitting $\rho \propto a^{-3-3w}$ for each stage of the evolution.  While this comparison is only strictly precise in the FLRW limit, the near-homogeneity of the lapse suggests that we can conceptually identify our $t = {\rm const}$ hypersurfaces as near-FLRW slices.

\Fig{fig:rho_vs_a} helps to contextualize how the fluid and field contribute to the overall dynamics of the box: the fluid (magenta) dilutes as radiation $\rho_{\rm fl} \propto a^{-4}$, as expected, and we compute its equation of state parameter to be $w \approx 0.33$ throughout the simulation. 
We now consider the field (cyan).
Between lines A and B, i.e. up to $t= 19.47\,m^{-1}$, the bubble is expanding but has not yet expanded fully across the grid, so some of the field remains in its false minimum. Thus, at this stage we find that the field
acts approximately as vacuum energy as the computed equation of state parameter $w\approx -0.96$ approaches $w = -1$.
In the next region between lines B and C for times $ 19.47\,m^{-1}-43.82\,m^{-1}$, the bubble walls have begun colliding; 
here we find $w \approx -0.62$, signaling that the field is transitioning away from vacuum-energy-like behavior.
As the bubble walls continue to collide and the field settles around its true minimum between C and D, times $43.82\,m^{-1}-116.9\,m^{-1}$, we find that the field dilutes like radiation with $w\approx0.31$.
The slight decrease in the computed equation of state parameter after line D ($t>116.9\,m^{-1}$), $w \approx 0.29$, may indicate the onset of matter-like behavior; we leave further discussion of the late stages of our simulations to \Sec{sec:oscillons}.

\subsection{Stochastic Gravitational Waves}\label{sec:GWs}

One consequence of cosmological FOPTs is the copious production of stochastic gravitational waves. A single, isolated bubble is spherically symmetric and does not radiate; gravitational waves are sourced once bubble walls collide and break that symmetry. On a periodic lattice, this includes collisions of a bubble with its own periodic images. In addition to the scalar field gradients themselves, other components of the Universe can also source gravitational radiation. When the plasma couples directly to the field, the expanding walls drive sound waves and turbulence that are often the dominant source~\cite{Hindmarsh:2013xza,Giblin:2013kea,Giblin:2014qia,Hindmarsh:2015qta}. In our setup, the radiation fluid responds to the bubbles only through gravity.

We understand gravitational waves as propagating transverse-traceless (TT) tensor perturbations of the metric, $g_{ij} = a^2(\delta_{ij} + h_{ij})$ with $\partial_i h_{ij} = h_{ii} = 0$. In an expanding background, the linearized Einstein equations give~\cite{Dufaux:2007pt}
\begin{equation}
    \ddot{h}_{ij} + 3H\dot{h}_{ij} - \frac{\nabla^2}{a^2}h_{ij} = \frac{16\pi}{\mpl^2 a^2}\, T^{\rm TT}_{ij},
\end{equation}
where dots denote derivatives with respect to cosmic time, $t$. The source is the TT part of the stress tensor, equivalently the TT part of the anisotropic stress, since the projection removes the trace. In Fourier space, $T^{\rm TT}_{ij}$ is obtained with the projector
\begin{equation}
    P_{ij} = \delta_{ij} - \frac{k_i k_j}{k^2},
\end{equation}
as
\begin{equation}\label{eq:TmnTT}
    T^{\rm TT}_{ij}(\vec{k}) = \sum_{lm}\left(P_{il}P_{jm} - \frac{1}{2} P_{ij}P_{lm}\right) T_{lm}(\vec{k}),
\end{equation}
where, at this linear order, spatial indices are raised and lowered with $\delta_{ij}$.

Given $h_{ij}(t,\vec{x})$, the effective stress-energy of the gravitational waves is~\cite{MisnerThorneWheeler1973}
\begin{equation}
    T^{\rm GW}_{\mu\nu} = \frac{\mpl^2}{32\pi}\left\langle \partial_\mu h_{ij}\, \partial_\nu h_{ij}\right\rangle,
\end{equation}
where $\langle\cdots\rangle$ denotes an average over many wavelengths. The energy density is therefore
\begin{equation}
    \rho_{\rm GW} = \frac{\mpl^2}{32\pi}\left\langle\dot{h}_{ij} \dot{h}_{ij}\right\rangle.
\end{equation}
Fourier transforming, the gravitational wave energy per logarithmic wavenumber, normalized to the critical density $\rho_c = 3H^2\mpl^2/8\pi$, is
\begin{equation}\label{eq:OmegaGW_hijdot}
    \Omega_{\rm GW}(k) = \frac{1}{\rho_c}\frac{d\rho_{\rm GW}}{d\ln k} = \frac{k^3}{24\pi^2 H^2 L^3}\sum_{ij}\left|\dot{h}_{ij}(t,\vec{k})\right|^2,
\end{equation}
where $L^3$ is the coordinate volume of the simulation, and the right-hand side is averaged over shells of constant $|\vec{k}|$. These expressions are strictly valid only for linear perturbations about an FLRW background. We use them here as a diagnostic of the gravitational radiation in our fully nonlinear simulations, in which the lapse remains close to unity, so coordinate time closely tracks cosmic time.

\begin{figure}
    \centering
    \includegraphics[width=0.99\linewidth]{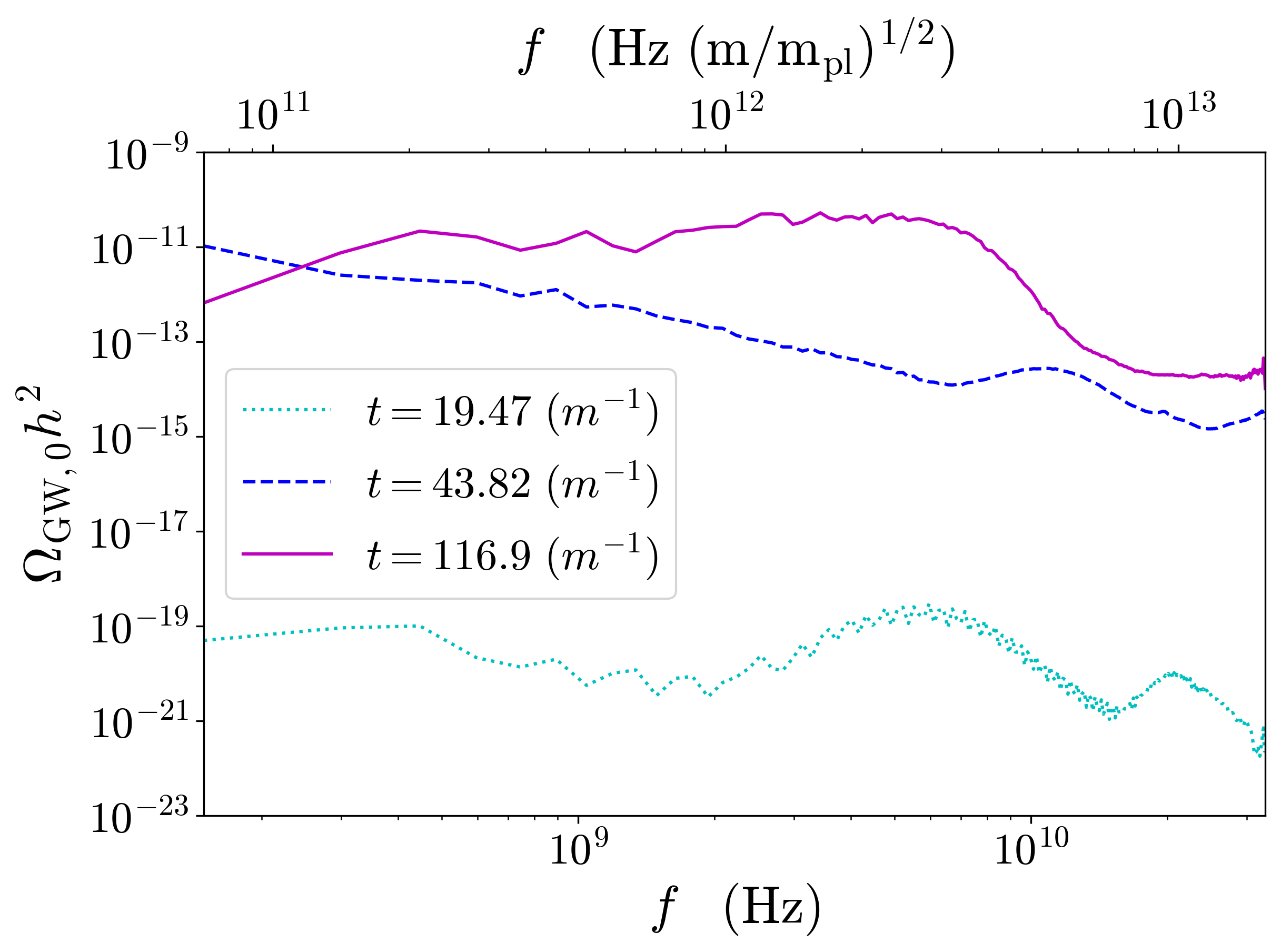}
    \caption{Gravitational wave power spectrum redshifted to today $\Omega_{\rm GW,0}h^2$ at times $t= 19.47\,m^{-1}$ (B; dotted cyan), $t=43.82\,m^{-1}$ (C; dashed blue), and $t=116.9\,m^{-1}$ (D; solid magenta) corresponding to the final three panels in \Fig{fig:phi_E_panels}. }
    \label{fig:OmegaGW_evolution}
\end{figure}

\begin{figure}
    \centering
    \includegraphics[width=0.99\linewidth]{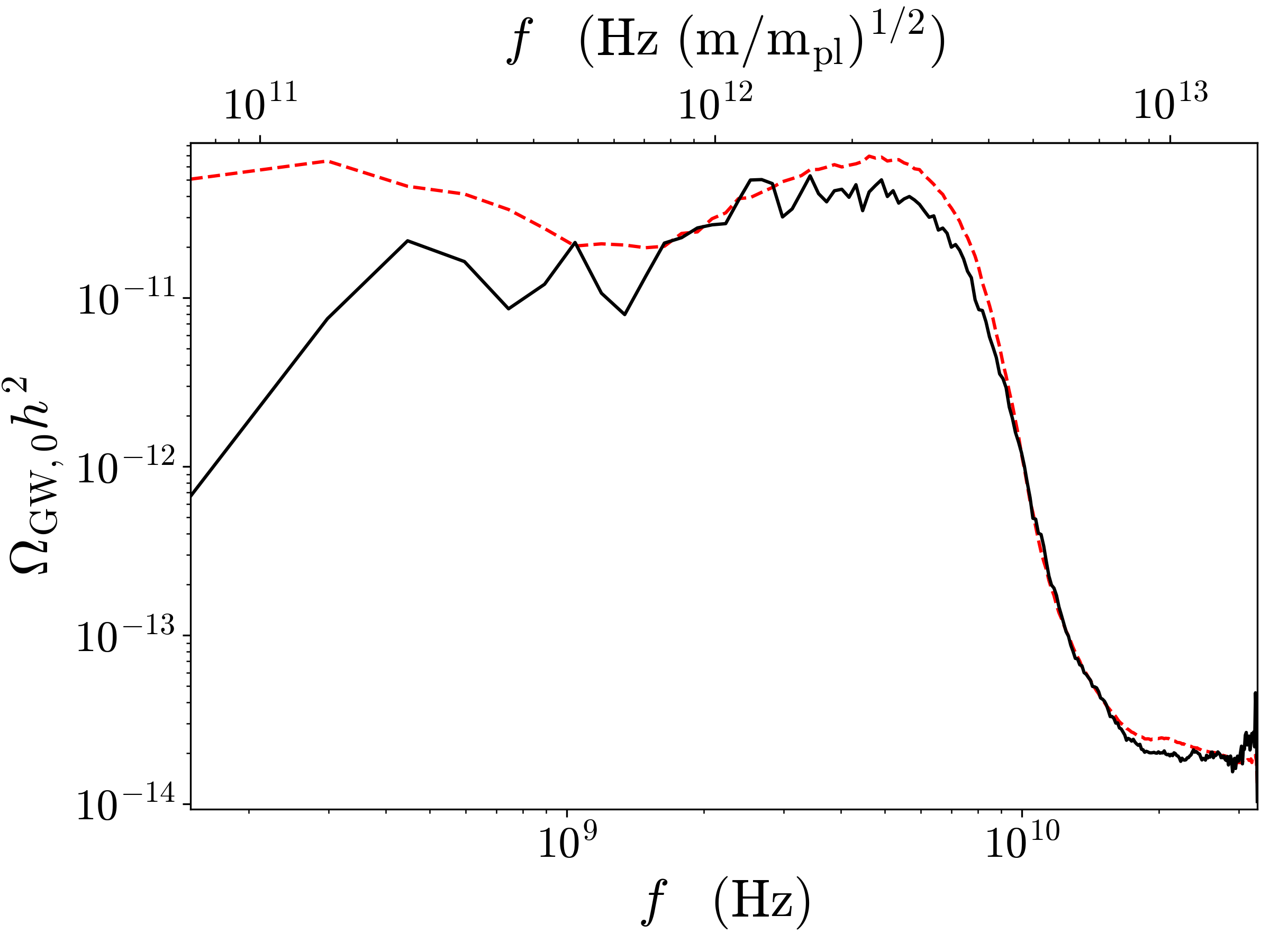}
    \caption{Gravitational wave power spectrum redshifted to today $\Omega_{\rm GW,0}h^2$ for a simulation with one bubble (solid black) versus two bubbles (dashed red) at $t=116.9\,m^{-1}$ (D).}
    \label{fig:OmegaGW_1v2bub}
\end{figure}

To compute the gravitational wave power spectrum in our simulations, we use the relationship $\dot{h}_{ij}\approx-2 \tilde{A}_{ij}$~\cite{Bastero-Gil:2010tpb}.
This uses $\alpha\approx1$ and neglects the shift terms, which remain small in our simulations. We can then rewrite \Eq{eq:OmegaGW_hijdot} in terms of $\tilde{A}_{ij}$, a \BSSN variable already computed during the evolution,
\begin{equation}\label{eq:OmegaGW_Aijsq}
    \Omega_{\rm GW}(k) = \frac{3k^3}{2\pi^2 \langle K\rangle^2 L^3}\sum_{ij}\left|\tilde{A}^{\rm TT}_{ij}(t,\vec{k})\right|^2,
\end{equation} 
where $L$ is the length of our box and $\langle K\rangle$ is the average of the trace of the extrinsic curvature related to the Hubble parameter as $\langle K\rangle = -3H$.
For more details on how $\Omega_{\rm GW}$ is computed in \texttt{GABERel}, please refer to \App{app:OmegaGW-in-GABERel}.

Finally, we redshift the spectrum to the present day. Assuming the Universe is radiation dominated from the end of our simulations until matter-radiation equality,
\begin{equation}\label{eq:OmegaGW_today}
    \Omega_{\rm GW,0}(k)h^2=\Omega_{\rm GW}(k)\left(\frac{g_0}{g_*}\right)^{1/3}\Omega_{r,0}h^2,
\end{equation}
where $h$ is the dimensionless Hubble parameter (not to be confused with $h_{ij}$), $g_*$ and $g_0$ are the effective numbers of relativistic degrees of freedom at the time the gravitational waves are produced and today, for which we take $g_*/g_0 \approx 100$, and $\Omega_{r,0}h^2\approx4.15\times 10^{-5}$ is the present radiation density. Similarly, the present-day frequency is
\begin{align}\label{eq:f_today}
    f &= \frac{k}{2\pi a\,\rho^{1/4}} \, \rho_{r,0}^{1/4}
    \left( \frac{g_*}{g_0} \right)^{1/4}
    \left( \frac{g_{S,*}}{g_{S,0}} \right)^{-1/3} \nonumber \\
    &\approx 3.2 \times 10^{10}\,\mathrm{Hz} \, \frac{k/a}{\sqrt{H \MPl}}
    \left( \frac{g_*/g_0}{100} \right)^{-1/12},
\end{align}
where $a$, $\rho$, and $H$ are evaluated at the time the spectrum is computed, $g_{S,*}$ and $g_{S,0}$ are the effective numbers of entropic degrees of freedom at production and today, and $\MPl$ is the reduced Planck mass. In the second line we set $g_S = g$. If the field's late-time, matter-like behavior (\Sec{sec:oscillons}) persisted long enough to dominate the expansion, these expressions would overestimate the present-day amplitude; here we assume the field decays into radiation before that happens.

\Fig{fig:OmegaGW_evolution} shows the present-day gravitational wave spectrum from our single-bubble simulation at $t= 19.47\,m^{-1}$ (B, dotted cyan), $t=43.82\,m^{-1}$ (C, dashed blue), and $t=116.9\,m^{-1}$ (D, solid magenta), the times of the final three panels of \Fig{fig:phi_E_panels}. Because the box is periodic, this simulation describes a cubic lattice of identical bubbles, and the radiation is sourced by collisions of the bubble with its periodic images. The bottom axis gives the frequency for our fiducial mass, $m = \sqrt{\lambda}\varphi_0 \approx 4.3\times10^{-6}\mpl$. The top axis gives the frequency for arbitrary $m$. In program units, $\lambda$ drops out of the equations of motion (see \App{appsec:UnitsAndRescaling}), so for fixed $\epsilon$, $\xi$, and $\varphi_0/\mpl$ the dynamics, and hence $\Omega_{\rm GW}$, are unchanged, while the present-day frequency scales as $f\propto\sqrt{m/\mpl}$~\cite{Giblin:2014gra}.

At $t= 19.47\,m^{-1}$, the bubble is expanding but its walls have not yet collided, and the spectrum remains small. By $t=43.82\,m^{-1}$, the walls have begun to collide, and the spectrum grows across all scales. By $t=116.9\,m^{-1}$, the field has settled into its true minimum, collisions have ceased, and the spectrum has reached its final form. It peaks at $\Omega_{\rm GW,0}h^2 \sim 5\times 10^{-11}$ near $f\approx 3-7\,{\rm GHz}$, with less power at higher frequencies. This amplitude is consistent with previous studies that do not include nonlinear gravity~\cite{Huber:2008hg,Child:2012qg,Cutting:2018tjt,Giblin:2014gra}.

For our fiducial parameters, the signal lies in the GHz band targeted by proposed high-frequency gravitational wave detectors~\cite{Aggarwal:2020olq}, although well below their projected sensitivities. Because the frequency scales with $\sqrt{m/\mpl}$, the same dimensionless dynamics at a transition scale of roughly $\mu = 10^{-16}\mpl\approx 1.22\,{\rm TeV}$ would place the peak at $3-7\,{\rm mHz}$. There it would lie above the projected LISA sensitivity~\cite{Caprini:2019egz,LISA:2024hlh}. We note that this rescaling requires a much smaller self-coupling, so it describes a transition with the same dynamics, not the same potential.

To compare simulations with different numbers of bubbles, we present the gravitational wave power spectrum redshifted to today for one (solid black) and two (dashed red) bubbles as the phase transition is completing at $t=116.9\,m^{-1}$ (D) in \Fig{fig:OmegaGW_1v2bub}. 
The two spectra display similar features; they peak around the same frequency and decay for higher frequencies.
The largest difference appears at low frequencies. At these scales, the spectrum is computed from very few modes, so we do not attribute the visual difference at low frequencies to any physical difference between the simulations. 
Overall, our results are consistent with Ref.~\cite{Child:2012qg}, where varying the number of bubbles per Hubble volume did not significantly change the amplitude of the gravitational wave spectrum. 

\begin{figure}
    \centering
    \includegraphics[width=0.99\linewidth]{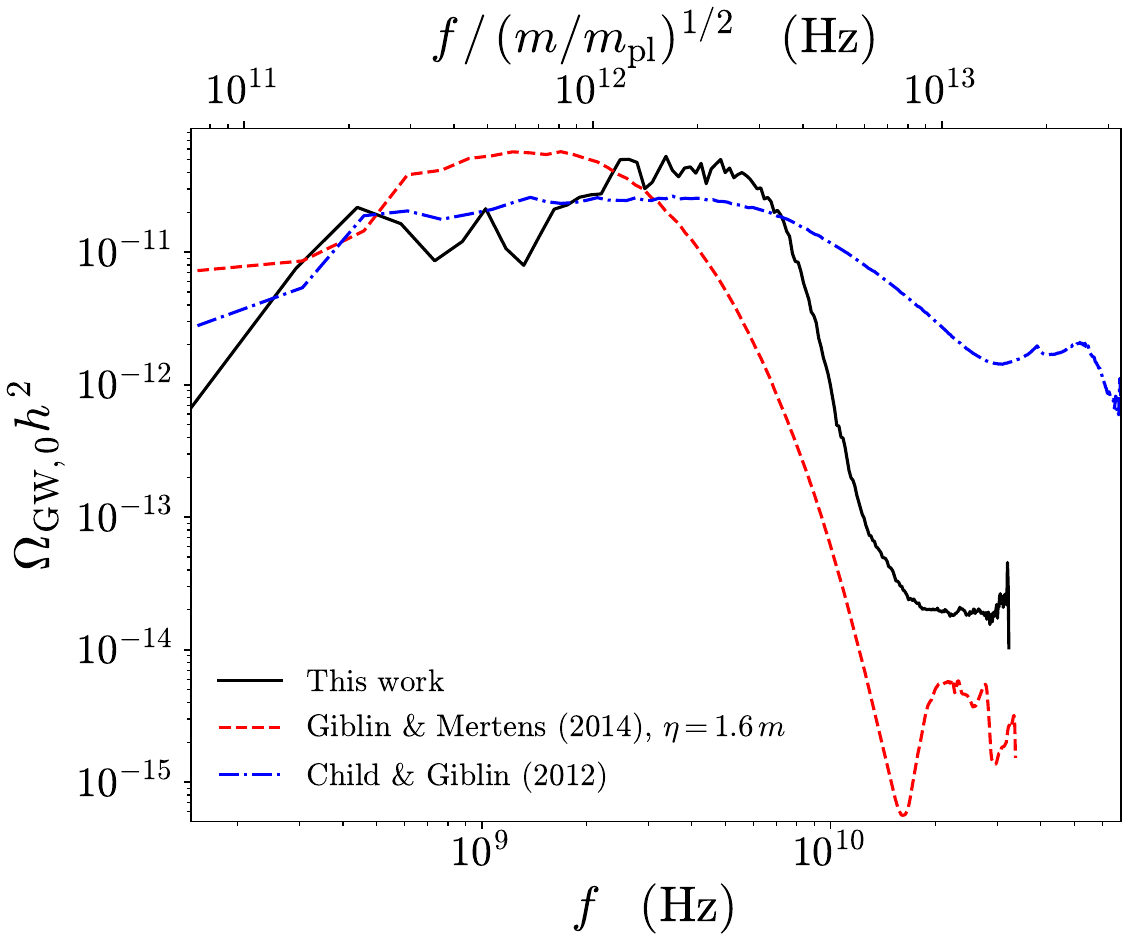}
    \caption{Present-day gravitational wave spectrum from our single-bubble simulation
at $t = 116.9\,m^{-1}$ (D; solid black) compared with two earlier lattice simulations
of the same type of transition that do not include nonlinear gravity.
The dashed red line is the simulation of Ref.~\cite{Giblin:2014qia} with transition strength $\alpha_{\rm GM} = 0.45$,
with 32 bubbles on a $256^3$ grid, a ratio of field to fluid energy of $\chi = 0.1$,
and a field--fluid coupling $\xi_{\rm GM} = 1.6$ ($\alpha_{\rm GM}$ and $\xi_{\rm GM}$ are defined in \cite{Giblin:2014qia}).
The dot-dashed blue line is the five-bubble simulation of Ref.~\cite{Child:2012qg},
with no fluid, $\xi = 0.25$ and $\mu = 10^{-4}\mpl$ on a $512^3$ grid,
shown once the scale factor has grown by $a = 1.5$.
All three simulations use a box of side $L = 1/(2H_*)$, and both earlier spectra
have been converted to the conventions of \Eqs{eq:OmegaGW_today}{eq:f_today}.
The features above $f \approx 2\times10^{10}\,{\rm Hz}$ in the earlier spectra are
numerical artifacts.}
    \label{fig:compare}
\end{figure}

We can also compare these simulations to some prior work that uses an FLRW background.  \Fig{fig:compare} compares the final
spectrum from our single-bubble simulation with two earlier lattice studies of
FOPTs that did not include nonlinear gravity. The
first~\cite{Child:2012qg} evolved the same potential, \Eq{eq:fullpot}, with five
bubbles in an expanding, radiation-dominated background, but without a fluid
and with a larger fraction of the energy in the field, $\xi = 0.25$. The
second~\cite{Giblin:2014qia} evolved 32 bubbles coupled to a relativistic fluid
through a friction term of strength $\eta$ (unrelated to the gauge parameter of \Sec{ssec:BSSNevolution}), with a ratio of field to fluid
energy $\chi = 0.1$; we show the case $\eta = 1.6\,m$. All three simulations use
a box of side length $L = 1/(2H_*)$. We have converted both earlier spectra to the
conventions of \Eqs{eq:OmegaGW_today}{eq:f_today}, and we place them at our fiducial
energy scale, $\mu = 10^{-4}\mpl$. Despite the differences in bubble
number, field--fluid coupling, energy fraction and treatment of gravity, the
three spectra peak in close proximity, at
$\Omega_{\rm GW,0}h^2 \approx (3-6)\times10^{-11}$. Our spectrum and that of
Ref.~\cite{Child:2012qg} also peak at nearly the same frequency,
$f \approx 3.5\,{\rm GHz}$, while the spectrum of Ref.~\cite{Giblin:2014qia} peaks
about a factor of two lower. The spectra differ more at high frequencies, where
ours falls off more steeply than that of Ref.~\cite{Child:2012qg}. The features
above $f \approx 2\times10^{10}\,{\rm Hz}$ in the earlier spectra are numerical
artifacts of those simulations. That a fully nonlinear simulation reproduces
the amplitude and peak frequency of these earlier results suggests that, for
transitions at this scale, nonlinear gravity does not substantially change the
gravitational wave signal near its peak.

\subsection{Late-Time Evolution and Oscillons}\label{sec:oscillons}
As the phase transition completes, the gravitational wave source vanishes and the spectrum reaches its final amplitude.
Although the bubble walls are no longer present and therefore do not lead to further collisions, the field remains dynamic as it oscillates about its true minimum. 
In these final stages, we find that the scalar field forms long-lived overdense regions, or oscillons~\cite{Bogolyubsky:1976yu,Gleiser:1993pt,Copeland:1995fq,Salmi:2012ta,Kou:2019bbc,Kou:2021bij,Nazari:2020fmk}.
Given the nonrelativistic nature of oscillons, they contribute to the energy density as matter. 
Consistent with this, after line D in \Fig{fig:rho_vs_a} the field's equation of state parameter decreases slightly, from $w\approx0.31$ to $w\approx0.29$, suggesting a growing matter-like contribution to the energy density of the field.

\begin{figure}
    \centering
    \includegraphics[width=0.99\linewidth]{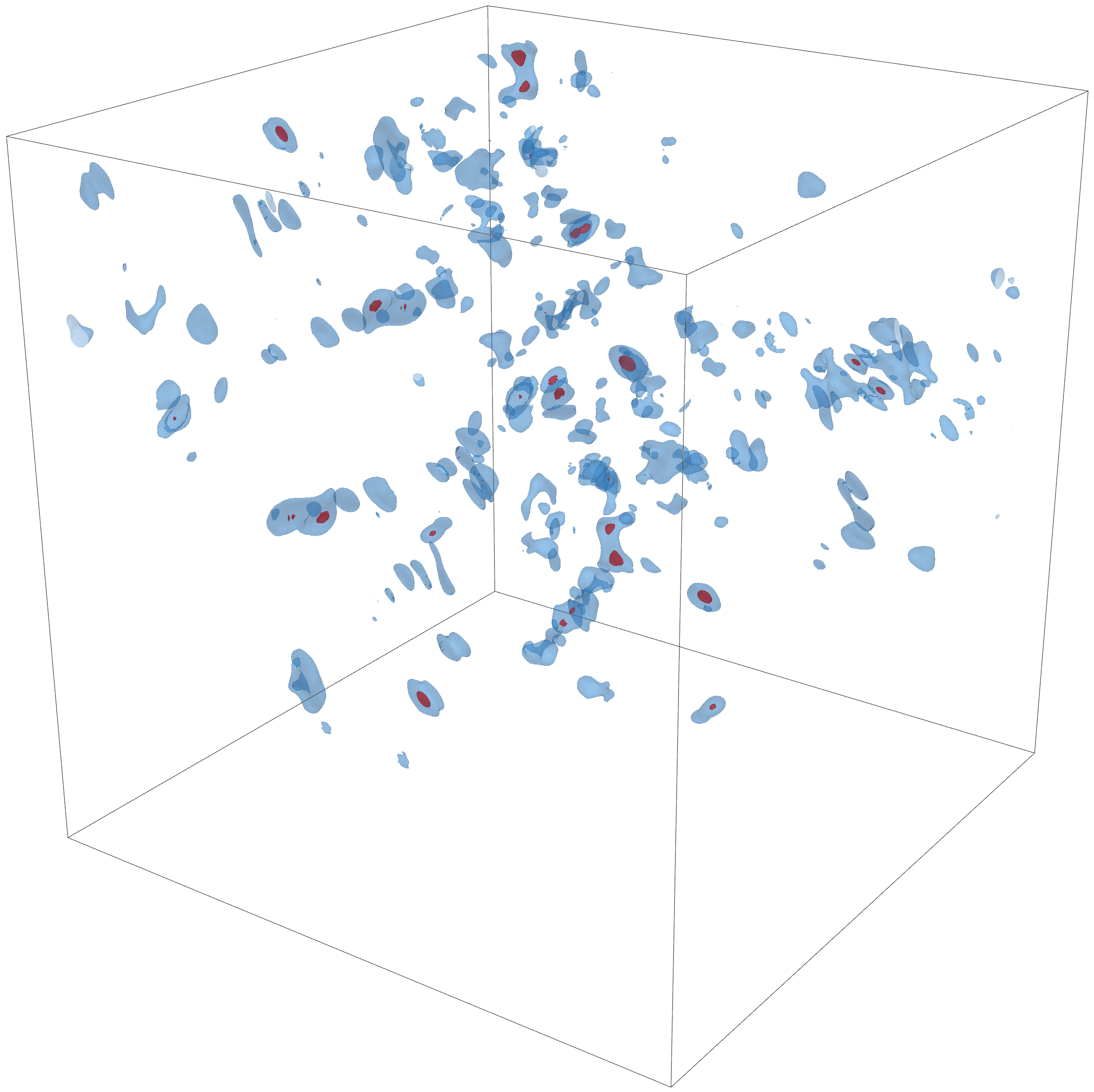}
    \caption{Energy density of the field normalized by the volume-averaged total energy density, $\rho_\varphi/\langle\rho\rangle$, from our two-bubble simulation as the phase transition is completing at time $t=116.9\,m^{-1}$.
    Blue contours indicate $\rho_\varphi=3\times\langle\rho\rangle$ and red contours indicate $\rho_\varphi=10\times\langle\rho\rangle$.
    }
    \label{fig:oscillons}
\end{figure}

We present the formation of oscillons from our simulation with two bubbles in \Fig{fig:oscillons}.
We note that the two-bubble simulation provides a better description of the oscillon distribution, as a second bubble breaks the cubic symmetry of the single-bubble lattice.
To show the oscillons, we present density contours to depict surfaces where regions of the field energy density $\rho_\varphi$ are much larger than the volume-averaged total energy density $\langle\rho\rangle$.
More specifically, blue contours indicate $\rho_\varphi = 3\times\langle\rho\rangle$ and red contours indicate even denser regions of $\rho_\varphi = 10\times\langle\rho\rangle$. 
We find that overdense regions of this size begin to form as the bubble walls collide 
and persist throughout the rest of the simulation.
In fact, we find that our bubble collisions form even larger overdense regions with a maximum of $\rho_\varphi \sim 100\times\langle\rho\rangle$ at time $t = 60.38\,m^{-1}$ when the final remnants of the bubble walls are colliding;
this moment in particular describes the highest energy collision in our simulation as the walls have had as much time as possible to gain momentum before colliding with each other.
Similarly, 
we note that the single-bubble simulation generates even larger overdense regions in the final stages of the bubble walls colliding as they have more time to expand before colliding in comparison to the two-bubble simulation. 
Regardless of the number of bubbles, we find that once the field has reached its true minimum, the overdense regions remain with a maximum of $\rho_\varphi \approx 10\langle\rho\rangle$. 

\section{Discussion}
\label{sec:discussion}

In this paper, we present simulations of a cosmological FOPT using fully nonlinear gravity in the presence of a radiation fluid.  We use \texttt{GABERel} to evolve the scalar field that undergoes the transition, the relativistic fluid, and the gravitational degrees of freedom in the \BSSN formalism on a periodic, Cartesian domain.  We focus first on a GUT-scale transition where the initial bubbles are comparable in size to the Hubble horizon.

As part of this setup, we developed a three-step algorithm to construct initial data containing bubbles.  This method first solves for an approximate field profile, as in Ref.~\cite{Coleman:1977py}, using a shooting method.  After this, we vary the bubble size to conserve energy during the nucleation event and keep the asymptotic expansion rate unchanged, while satisfying the Hamiltonian constraint, \Eq{eq:hamconst} (and trivially the momentum constraint, \Eq{eq:momconst}), on the initial surface.  The resulting profile agrees with the thin-wall approximation to within $\sim 2\%$ in the bubble walls.

We then study the phase transition as it evolves: the bubbles expand, collide, and coalesce.  We confirm that the radiation fluid is churned during the phase transition with percent-level density contrasts, $|E-\langle E\rangle|/\langle E \rangle \lesssim 0.02$.  The field, on the other hand, dilutes first as a vacuum-energy-like component until the majority of the volume is in the true minimum.  The field then dilutes like radiation and finally as a combination of radiation and matter.  We also compute the stochastic gravitational waves from the trace-free part of the conformal extrinsic curvature, $\tilde{A}_{ij}$.  This gravitational wave signature peaks at $\Omega_{\rm GW,0}h^2 \sim 5\times10^{-11}$, which is consistent with prior work~\cite{Huber:2008hg,Child:2012qg,Cutting:2018tjt,Giblin:2014qia,Giblin:2014gra}.  We also note that the late-time, post-transition state appears to indicate the presence of oscillons, pushing the field's equation of state toward that of matter.

The simulations here present a first pass at examining FOPTs in full numerical relativity.  We have limited our simulations to one- and two-bubble scenarios, which ties the duration of the simulation to the box size.  While both of these quantities are chosen so that they are related to the Hubble scale, more simulations with different configurations of bubbles can be helpful to ensure the robustness of the predicted gravitational wave signal.  In the simulations presented here, we see a final state where the variation of the lapse and shift indicates that we end on a near-FLRW foliation; we plan to extend the parameter space of these simulations to see if FOPTs can end in a less homogeneous final state.  If such scenarios can become significantly more violent, this approach would allow us to understand whether black holes can form during these transitions~\cite{Hawking:1982ga,Kodama:1982sf,Jung:2021mku,Liu:2021svg,Baker:2021nyl,Kawana:2021tde, Jinno:2023vnr}.

\acknowledgments
We want to thank Thomas Baumgarte, Sam Brady, Katy Clough, Anne-Christine Davis, Dani Figueroa, Mark Hindmarsh, Eugene Lim, Avery Tishue, and Areef Waeming for discussions that helped shape the direction of this work. We are grateful for inspiration from our earlier work with Hillary Child, Jim Mertens and Eva Nesbit.

C.B., J.T.G. and C.R. are supported by the National Science Foundation, PHY-2309919, awarded to Kenyon College.  Simulations for this work were carried out on hardware purchased through this award, and previous support from the National Science Foundation and Kenyon College.


\appendix 

\section{Units and Rescaling}\label{appsec:UnitsAndRescaling}

In addition to decomposing our four-dimensional system and recasting the $3+1$ constraint and evolution equations into the \BSSN formulation, we also introduce a rescaling of our parameters.
To express our variables in program units, denoted by subscript ``pr'', we introduce a rescaling factor 
\begin{equation}
    B = \frac{\sqrt{\lambda}\varphi^2_0}{\mpl} = m\frac{\varphi_0}{\mpl},
\end{equation}
so that the program potential takes the form 
\begin{equation}
    V_{\rm pr} = \frac{V}{B^2\mpl^2}=\frac{1}{8}\left(\frac{\varphi^2}{\varphi^2_0}-1\right)^2+\epsilon\left(\frac{\varphi}{\varphi_0}+1\right) + \frac{V_0}{\lambda\varphi_0^4}.
\end{equation}
Under this rescaling, $\lambda$ drops out of the equations of motion. For example, the Friedmann equation becomes $H_{\rm pr}^2 = 8\pi\rho_{\rm pr}/3$, and the evolution of $\varphi/\varphi_0$ depends only on $\epsilon$ and $\varphi_0/\mpl$. Dimensionless combinations such as $LH = L_{\rm pr}H_{\rm pr}$ are unchanged.

Given this rescaling, our length, time, and energy density in program units become
\begin{align}\label{eq:xpr}
    x_{\rm pr} &= xB, \\
    t_{\rm pr} &= tB, \\
    \rho_{\rm pr} &= \frac{\rho}{B^2 \mpl^2}=\frac{\rho}{\lambda \varphi_0^4},
\end{align}
and we identify the program initial bubble radius as
\begin{equation}\label{eq:rbub-pr}
    r_{*,\rm pr} = r_* B = \frac{\varphi_0}{\epsilon\mpl}.
\end{equation}


\section{Computing $\Omega_{\rm GW}$ in \texttt{GABERel}}\label{app:OmegaGW-in-GABERel}

In practice, computing the transverse-traceless components of $\tilde{A}_{ij}$ is rather costly.
Instead, we compute the right-handed ($+$) and left-handed ($-$) helicity polarizations of $\tilde{A}_{ij}$ in \texttt{GABERel},
as these are the only degrees of freedom in the TT gauge. 
The polarizations $A_+$ and $A_-$ are split into their real and imaginary parts, $A_\pm = \mathrm{Re}(A_\pm)+i\,\mathrm{Im}(A_\pm)$. 
They are then constructed from a combination of circular basis vectors,
\begin{align}
    \mathrm{Re}(\boldsymbol{e}_+)&=\frac{1}{\sqrt{2}\kappa}(-k_y,k_x,0)=\mathrm{Re}(\boldsymbol{e}_-),\\
    \mathrm{Im}(\boldsymbol{e}_+)&=\frac{1}{\sqrt{2}\kappa|\vec{k}|}(-k_x k_z,-k_yk_z,\kappa^2)=-\mathrm{Im}(\boldsymbol{e}_-),
\end{align}
where $\kappa \equiv \sqrt{k_x^2+k_y^2}$.
Both $\mathrm{Re}(\boldsymbol{e}_\pm)$ and $\mathrm{Im}(\boldsymbol{e}_\pm)$ are orthogonal to $\hat{k}$ and to each other, so that $\boldsymbol{e}_\pm\cdot\boldsymbol{e}_\pm^* = 1$.
The helicity components are obtained by contracting with the conjugate basis vectors,
\begin{equation}
    A_\pm \equiv \sum_{i,j=1}^{3} e^{*\,i}_\pm e^{*\,j}_\pm\,\tilde{A}_{ij}.
\end{equation}
Writing $\boldsymbol{e}^R_\pm \equiv \mathrm{Re}(\boldsymbol{e}_\pm)$ and $\boldsymbol{e}^I_\pm \equiv \mathrm{Im}(\boldsymbol{e}_\pm)$, and recalling that $\tilde{A}_{ij}(\vec{k})$ is complex, we find
\begin{align}
    \mathrm{Re}(A_\pm) &= \sum_{i,j=1}^{3} \Big[\left(e^{R\,i}_\pm e^{R\,j}_\pm-e^{I\,i}_\pm e^{I\,j}_\pm\right)\mathrm{Re}(\tilde{A}_{ij}) \nonumber \\
    &\qquad\quad + \left(e^{R\,i}_\pm e^{I\,j}_\pm+e^{I\,i}_\pm e^{R\,j}_\pm\right)\mathrm{Im}(\tilde{A}_{ij})\Big],\\
    \mathrm{Im}(A_\pm) &= \sum_{i,j=1}^{3} \Big[\left(e^{R\,i}_\pm e^{R\,j}_\pm-e^{I\,i}_\pm e^{I\,j}_\pm\right)\mathrm{Im}(\tilde{A}_{ij}) \nonumber \\
    &\qquad\quad - \left(e^{R\,i}_\pm e^{I\,j}_\pm+e^{I\,i}_\pm e^{R\,j}_\pm\right)\mathrm{Re}(\tilde{A}_{ij})\Big].
\end{align}

Combining this all together, we find the dimensionless gravitational wave power spectrum computed in our simulations to be
\begin{equation}\label{eq:OmegaGW}
    \Omega_{\rm GW}(k) = \frac{3\,\Delta x^6 k^3}{2\pi^2L^3\langle K\rangle^2}\sum_{\lambda=\pm}\left|A_{\lambda}\right|^2,
\end{equation}
where $\Delta x=L/N$ is the grid spacing, and the right-hand side is averaged over shells of constant $|\vec{k}|$; the factor $\Delta x^6$ converts the discrete Fourier transform to the continuum normalization of \Eq{eq:OmegaGW_Aijsq}.

\section{Dissipation}
\label{app:dissipation}

Given the discretization of our numerical grid, high-frequency modes with wavelengths similar to the grid spacing can 
generate numerical noise.
To address the noise from these modes (large $k$), we implement artificial dissipation, namely Kreiss--Oliger dissipation~\cite{KreissOliger1973}. 
We note that damping these modes does not impact the physics of our simulations since these modes are already under-resolved. 
Furthermore, without sufficient dissipation, we find that our simulations become unstable and either (i) develop sharp features that eventually lead to the simulation crashing or (ii) noise from higher frequencies propagates ``inward'' and contaminates lower frequencies.

In order to address the numerical noise while leaving the physics as unaffected as possible, we add dissipation to our metric and field quantities but do not add dissipation to our fluid variables as applying dissipation to the fluid can have a smearing effect.
To confirm the impact of dissipation on the fluid, we ran simulations with and without dissipation applied to the fluid variables $E$ and $P^i$; we find excellent agreement for the final gravitational wave power spectrum between these cases and determine that dissipation added to the fluid is not necessary to achieve numerical stability. 

\begin{figure*}
    \centering
    \includegraphics[width=0.325\linewidth]{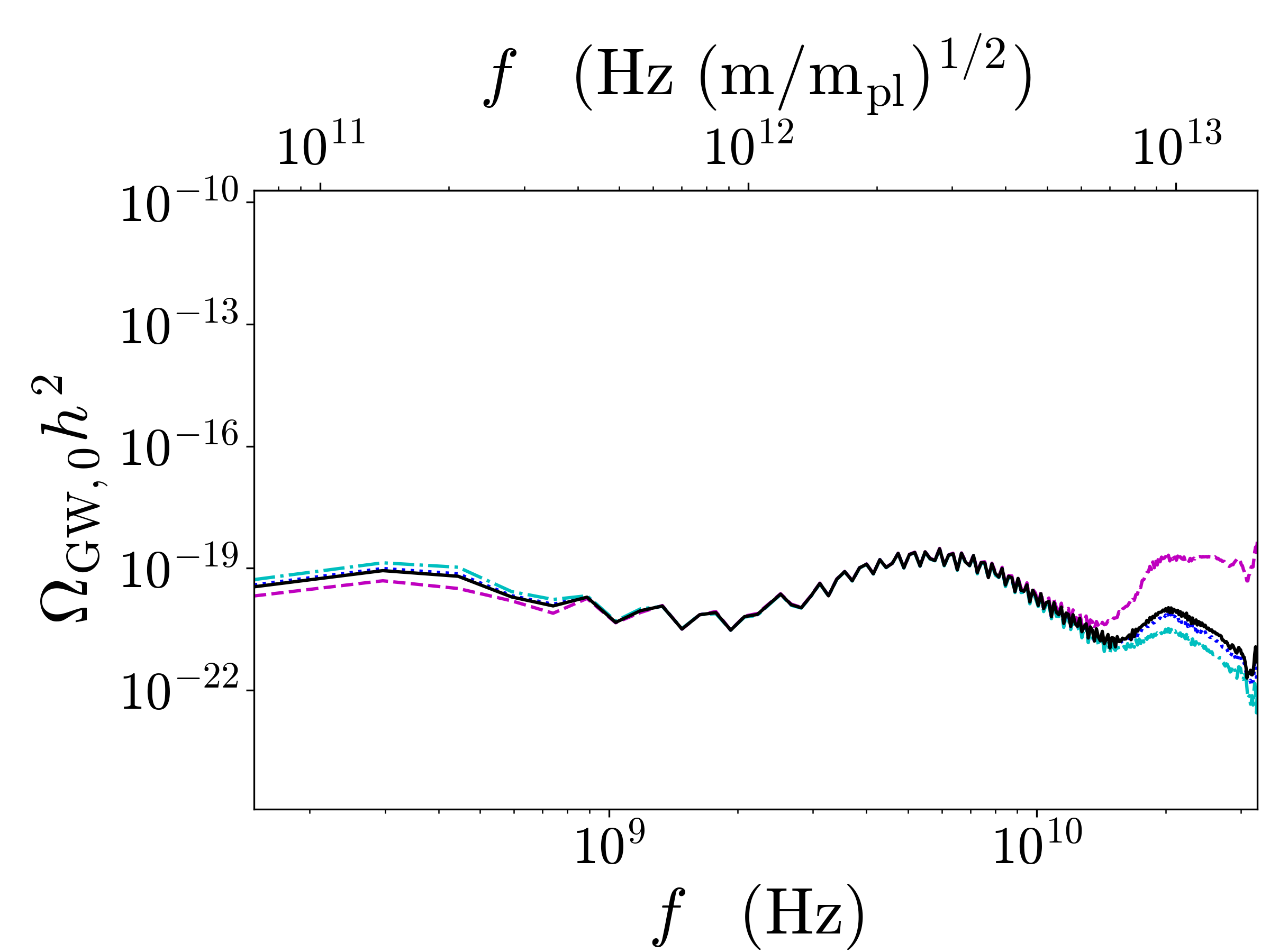}
    \includegraphics[width=0.325\linewidth]{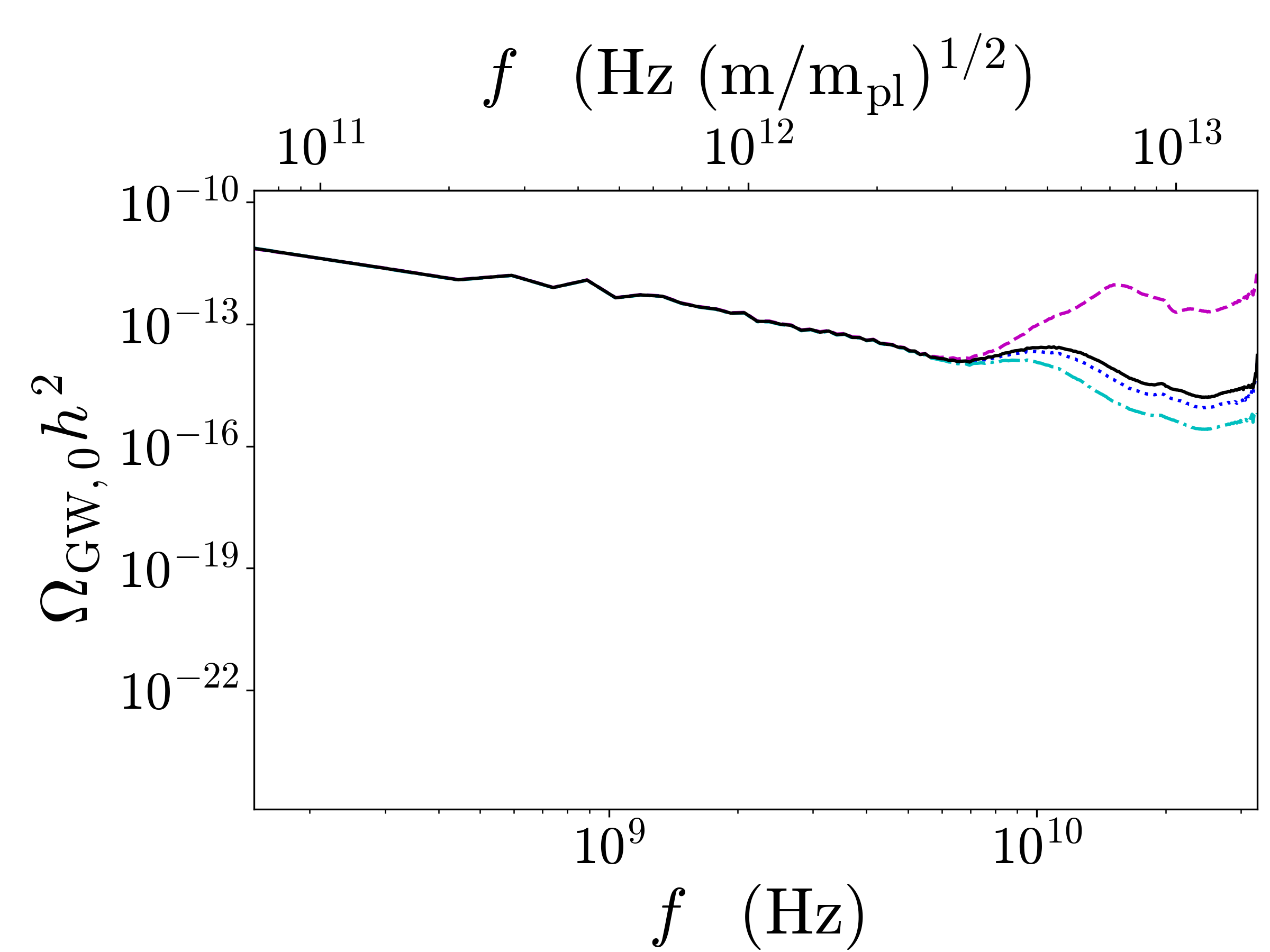}
    \includegraphics[width=0.325\linewidth]{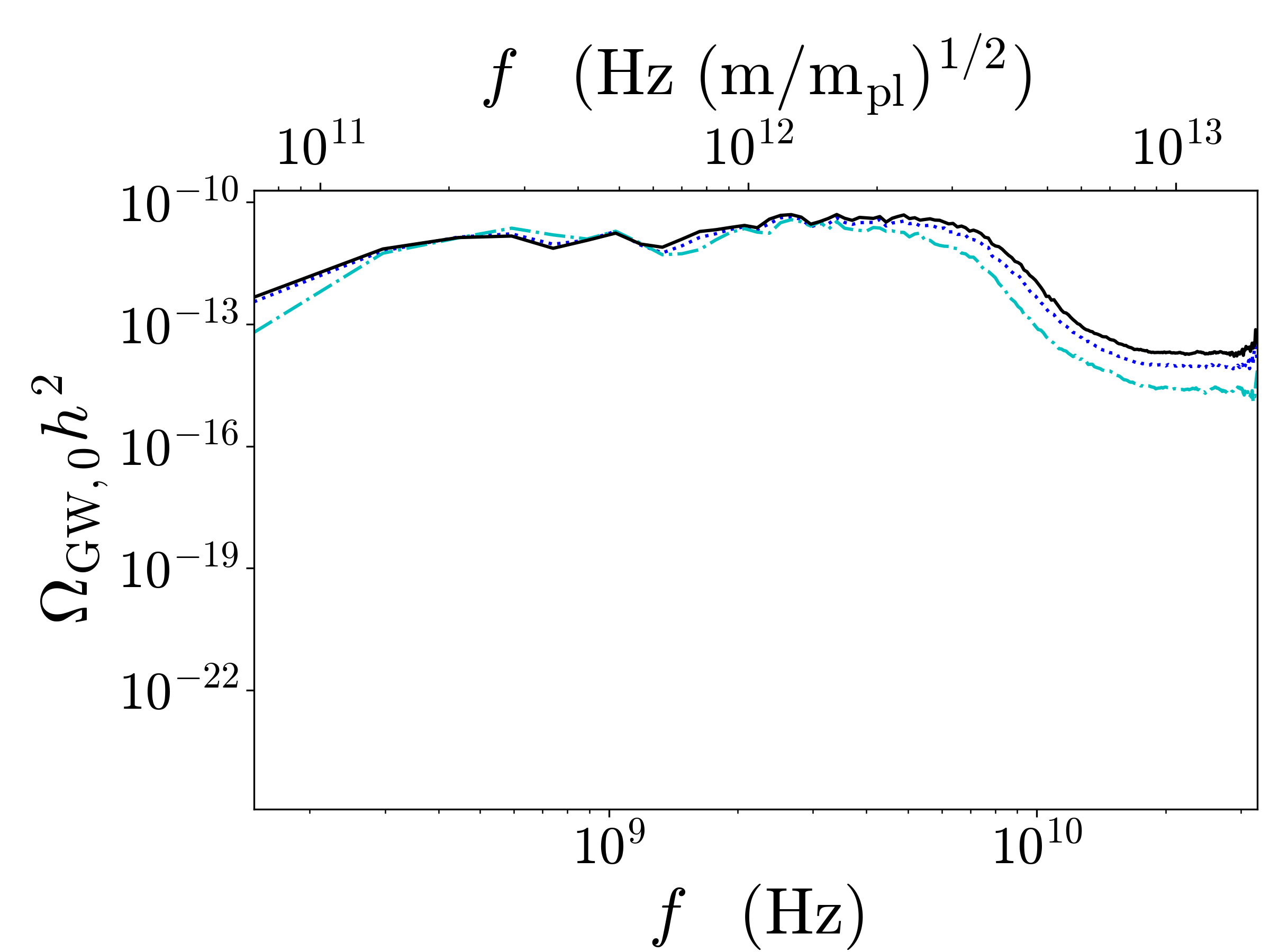}
    \caption{Comparison of the gravitational wave power spectrum redshifted to today, $\Omega_{\rm GW,0}h^2$, for varied dissipation parameters $\nu_{\rm dis} = 0$ (dashed magenta), $\nu_{\rm dis} = 0.4$ (solid black), $\nu_{\rm dis} = 0.5$ (dotted blue), and $\nu_{\rm dis} = 0.8$ (dash-dot cyan) at times $t= 19.47\,m^{-1}$ (B; left), $t=43.82\,m^{-1}$ (C; middle), and $t=116.9\,m^{-1}$ (D; right).
    }
    \label{appfig:omegaGW_comp_dis}
\end{figure*}

From von Neumann stability, we determine that the scheme is stable for choices of $\nu_{\rm dis}$ that satisfy Eq.~(9.9.7) of Ref.~\cite{alcubierre2008introduction},
\begin{align}
    0\leq \nu_{\rm dis} \frac{\Delta t}{\Delta x}\leq \frac{1}{2^{r-1}},
\end{align}
where $2r$ is the order of the Kreiss--Oliger derivative operator, with $r=3$ (sixth order) here, and $\Delta t/\Delta x = 1/30$ in our simulations, so all values considered here satisfy this bound. 
As a guide to understand which dissipation parameter is best suited to damp away numerical noise while leaving the physics unaffected, we present a comparison for the dimensionless gravitational wave power spectrum $\Omega_{\rm GW,0}h^2$ for dissipation parameters $\nu_{\rm dis} = 0$ (dashed magenta), $\nu_{\rm dis} = 0.4$ (solid black), $\nu_{\rm dis} = 0.5$ (dotted blue), and $\nu_{\rm dis} = 0.8$ (dash-dot cyan) at times $t= 19.47\,m^{-1}$ (B; left), $t=43.82\,m^{-1}$ (C; middle), and $t=116.9\,m^{-1}$ (D; right) in \Fig{appfig:omegaGW_comp_dis}.
For reference, the simulation with $\nu_{\rm dis} = 0.4$ (solid black) represents the single-bubble simulation presented in \Sec{sec:Results}. 

The comparison between the varied dissipation parameters in \Fig{appfig:omegaGW_comp_dis} highlights several features.
First, the simulation without dissipation (dashed magenta) does not appear in the third panel because the simulation becomes numerically unstable before the phase transition is complete.
This indicates that numerical dissipation is required to achieve numerical stability in our simulations. 
Next, we find that for dissipation parameters $\nu_{\rm dis} > 0$, the spectra in the first two panels agree for frequencies $f \lesssim 7\times10^9\,{\rm Hz}$ (i.e., below $\sim30\%$ of the one-dimensional Nyquist frequency);
we do not trust frequencies beyond this cutoff. 
By the third panel, the spectra separate slightly from each other, but the overall features of the spectrum are robust. 
Namely, most of the power remains in intermediate modes and less power is in higher-frequency modes. 
As a result, we choose the smallest nonzero dissipation parameter considered, $\nu_{\rm dis} =0.4$, for our production simulations because it leads to numerically stable simulations and maintains the features of the gravitational wave power spectrum without damping away physical signals.

\bibliography{Bibliography}

\end{document}